\documentclass[fleqn,usenatbib]{rasti}

\usepackage{newtxtext,newtxmath}

\usepackage[T1]{fontenc}

\DeclareRobustCommand{\VAN}[3]{#2}
\let\VANthebibliography\thebibliography
\def\thebibliography{\DeclareRobustCommand{\VAN}[3]{##3}\VANthebibliography}

\usepackage{graphicx}	
\usepackage{amsmath} 	

\title[\texttt{MHSXtraPy}]{\texttt{MHSXtraPy} - A Python code for the extrapolation of magnetohydrostatic fields on the Sun using analytical solutions}

\author[Nadol \& Neukirch]{
L. Nadol,$^{1}$\thanks{E-mail: lmn6@st-andrews.ac.uk}
T. Neukirch,$^{1}$
\\
$^{1}$School of Mathematics and Statistics, University of St Andrews, St Andrews, KY16 9SS, United Kingdom
}

\date{Accepted XXX. Received YYY; in original form ZZZ}

\pubyear{2025}

\begin{document}
\label{firstpage}
\pagerange{\pageref{firstpage}--\pageref{lastpage}}
\maketitle

\begin{abstract}
We present a Python code for calculating and displaying magnetic field extrapolations from given two-dimensional boundary conditions, specifically from solar surface magnetograms. The code implements analytical magnetohydrostatic models that incorporate the transition from non-force-free to force-free magnetic fields in the solar atmosphere. It allows for different parameterisations of this transition and includes functions to compute magnetic fields, plasma pressure, and density. Fast Fourier methods ensure efficient computation, and the output includes three-dimensional visualisations of field lines and plasma structures. The implementation is optimised for accessibility and speed, making it suitable for both research and educational purposes. The only prerequisite for running the code is a Python compiler. All source code, examples, input files, solutions, and instructions are available for download from GitHub.
\end{abstract}




\section{Introduction}

Magnetic field extrapolation methods based on photospheric observations as boundary conditions are an essential tool for investigations of the structure of the solar corona and the phenomena occurring within it.

Solar magnetic fields are categorised as force-free or as non-force-free depending on the ratio of plasma to magnetic pressure, also referred to as the plasma beta, $\beta_P$. As a consequence the magnetic fields in the lower layers of the solar atmosphere are regarded as non-force-free, while the coronal magnetic field is considered force-free. A transition between the two states takes place with increasing height; this change occurs mostly localised around a height of approximately 2 Mm above the photosphere \citep{Gary2001}, a region referred to as the transition region. 

However, force-free models are frequently used for the whole solar atmosphere as they are easy to apply and rely on a reduced set of equations. These models are categorised into potential, linear, and non-linear fields depending on the amplitude and spatial variability of the current density. While potential and linear models are commonly used as quick-look approaches, non-linear force-free (NLFF) methods  can capture certain aspects of the physics occurring on the Sun more accurately, e.g. localised twist and shear in field lines \citep[][]{Wiegelmann2017, WiegelmannSakurai2021}. 

Both linear and non-linear force-free models neglect plasma forces and assume that the Lorentz force vanishes only allowing currents parallel to the magnetic field. Therefore, a disadvantage of force-free methods remains the inconsistency between the assumption of a force-free photosphere in these model and the fact that the observed photospheric boundary conditions are not force-free (based on the definition of $\beta_P$). An alternative to force-free models is given by magnetohydrostatic (MHS) methods which additionally include perpendicular currents and as a consequence can capture the effects of pressure gradient and gravitational forces \citep{Zhu2022}.


By making additional assumptions, the MHS equations can lead to linear equations for which solutions can be obtained analytically without the need for numerical solvers. While such assumptions may limit the physical accuracy of the model, especially when it comes to representing twist in magnetic field lines, linear equations can be solved at much lower computational cost and therefore can be used as a complementary method to computationally expensive non-linear models. This is an advantage as recent improvements of observations and data availability lead to the need for magnetic field extrapolation models of ever higher resolution and run-time efficiency.  

Analytical MHS models have been around for decades \citep[e.g.][]{Low1991, Low1992} and have been continuously advanced to include more flexibility when modelling currents perpendicular to the magnetic field and to increase computational performance \citep[e.g.][]{Neukirch2019, Nadol2025}.  

This paper presents the \textbf{M}agneto\textbf{H}ydro\textbf{S}tatic e\textbf{Xtra}polation \textbf{Py}thon tool (short: MHSXtraPy), a code that was written specifically for the application of these analytical models. It provides an accessible, open-source tool that can be used for time-efficient investigations and visualisations of MHS fields from photosphere to corona. The language chosen is \texttt{Python} as it provides an ecosystem of libraries useful for our purposes. 

In the following we first introduce the theory behind the tool focusing on the output quantities of the code; second, we demonstrate its functionality using a SHARP magnetogram \citep{SHARP}; and last, briefly discuss technical aspects of the implementation and conclusions.

\section{MHSXtraPy}

\begin{figure*} 
 \centering
 \vspace*{1cm}
 \hspace*{2.8cm}
     \includegraphics[width=0.6\textwidth]{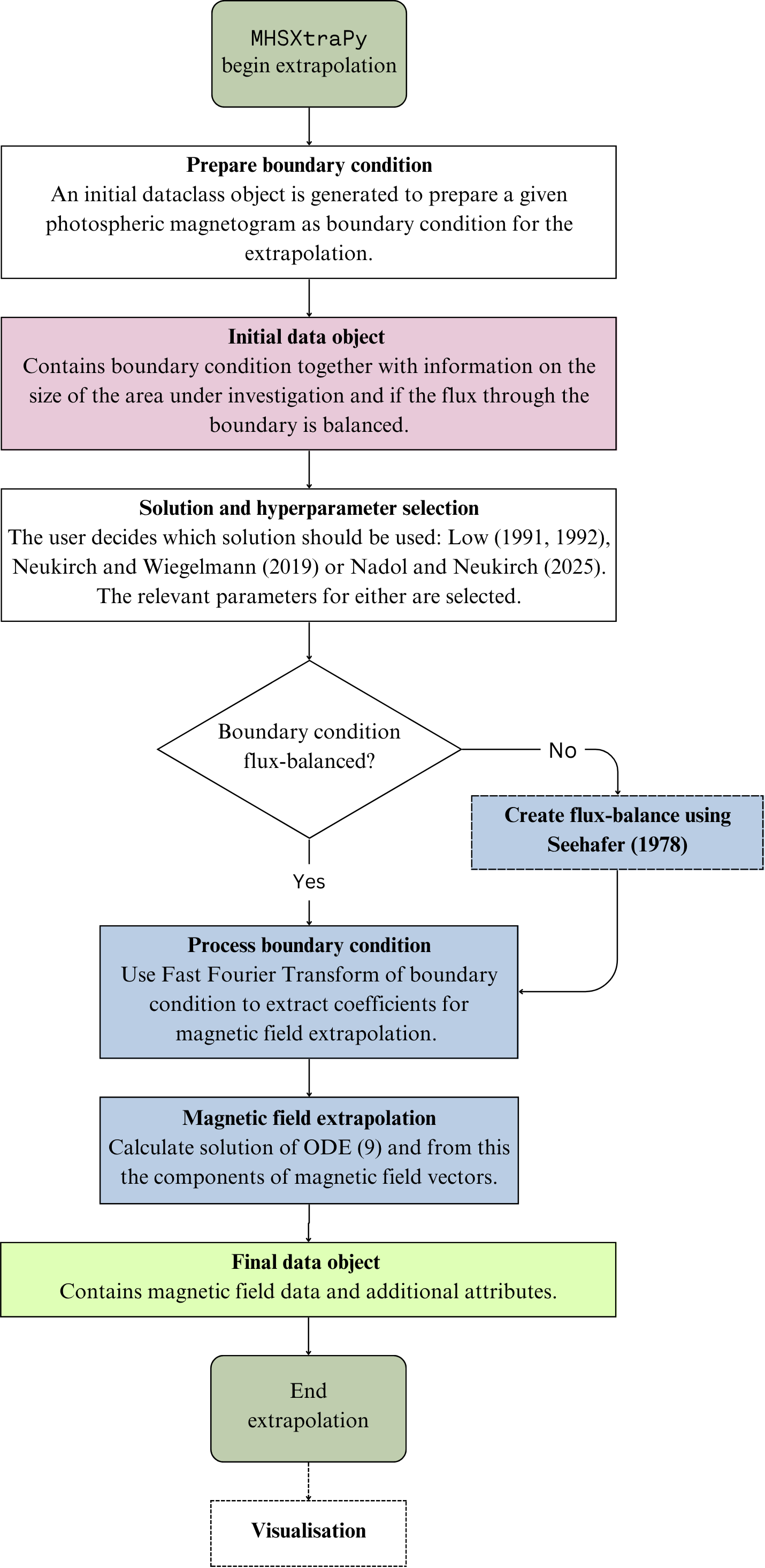}
     \caption{Flow chart representing the application of \texttt{MHSXtraPy}.} 
     \label{fig:flow}
 \end{figure*}

This section gives a brief overview of the background theory used for \texttt{MHSXtraPy}. How this connects to and is represented in the application of the code is shown in figure \ref{fig:flow}.

\subsection{Formulation}

In the MHS case the system of differential equations 
\begin{align}
\textbf{j} \times \textbf{B} - \nabla p - \rho g \hat{\textbf{z}} &= 0, \label{eq:momentum}\\
\nabla \times \textbf{B} &= \mu_0 \textbf{j}, \label{eq:curlB}\\
\nabla \cdot \textbf{B} &= 0 \label{eq:solenoidal},
\end{align}
where $\mu_0 = 4 \pi \times 10^-7$ Hm$^{-1}$ the permeability of a vacuum and $g$ the constant gravitational acceleration ($g = 272.2$ ms$^{-2}$ on the Sun), needs to be solved for the magnetic field vector \textbf{B}, the current density \textbf{j} as well as the plasma pressure $p$ and plasma density $\rho$ in combination with the ideal gas law
\begin{equation*}
	p = \rho RT / \bar{\mu},
\end{equation*}
where $R = 8 \times 10^3$ JK$^{-1}$kg$^{-1}$ the gas constant, $T$ the temperature and $\bar\mu = 1.67262 \times 10^{-7}$ kg the mean molecular weight. These are the four quantities \texttt{MHSXtraPy} determines. 

\subsubsection{Current Density}

In order for an analytical solution of Equations \eqref{eq:momentum}, \eqref{eq:curlB} and \eqref{eq:solenoidal} to exist, a specific form of the current density is used,
\begin{equation} 
\mu_0 \textbf{j} = \alpha \textbf{B} + f(z) \nabla B_z \times \hat{\textbf{z}}, \label{eq:j}
\end{equation}
where the function $f(z)$ governs the amplitude of the perpendicular currents in the system and $\alpha$ is the ratio of the toroidal to the poloidal components of $\textbf{B}$ (see equation \eqref{eq:poltor}). This function is therefore responsible for describing the transition from non-force-free to force-free magnetic fields in the equations of the model. 

Originally an exponentially decaying amplitude of the form
\begin{equation}
	f_L(z) = a_L \exp(-\kappa z), \label{eq:fLow}
\end{equation}
was suggested by \citet{Low1991, Low1992}, where $a_L$ is an amplitude parameter for the degree of non-force-freeness on the photosphere and $\kappa$ determines how fast the perpendicular currents drop off in the upper solar atmosphere, as $f_L(z) \to 0$ for $z \to \infty$. This solution has been used for extrapolation in \citet{Wiegelmann2015, Wiegelmann2017b}. 

Alternatively, \citet{Neukirch2019} suggested 
\begin{equation}
	f_{N+W}(z) = a_{N+W} \left[1-b\tanh \left(\frac{z-z_0}{\Delta z}\right) \right] \label{eq:fNW}
\end{equation}
to control the amplitude of the perpendicular currents with height $z$. This choice of $f(z)$ in equation \eqref{eq:j} increases the parameter space, and as a consequence the flexibility one has in modelling the transition from non-force-free to force-free. Similar to $a_L$ in equation \eqref{eq:fLow} the amplitude of $f_{N+W}$ is controlled by $a_{N+W}$. The additional parameters $z_0$ and $\Delta z$ determine the height at which that transition occurs and the width over which it takes place, respectively. The parameter $b$ can be interpreted as a "switch-off"-parameter as $b=1$ corresponds to a fully force-free state being reached above $z_0$; $b \neq 1$, on the other hand, allows for a certain degree of non-force-freeness being maintained in the upper atmospheric layers \citep[for details see][]{Neukirch2019, Nadol2025}.

Both of the above options lead to analytical expressions for the magnetic field \textbf{B}. Once \textbf{B} is calculated, \textbf{j} can be obtained using equation \eqref{eq:j}. 

\subsubsection{Magnetic Field}

Using the poloidal-toroidal representation \citep[e.g.][]{NeukirchRastst1999}
\begin{equation}
\textbf{B} = \nabla \times \left[ \nabla \times \left( \Phi \hat{\textbf{z}} \right) \right] + \nabla \times \left( \alpha \Phi \hat{\textbf{z}} \right) \label{eq:poltor}
\end{equation}
of the magnetic field in combination with a current density of form \eqref{eq:j} one can obtain a solution defined by
\begin{equation} \label{eq:intPhi}
\Phi(x, y, z) = \iint_{- \infty}^\infty \bar{\Phi}(z; k_x, k_y) \exp \left[ i \left(k_x x + k_y y \right) \right] dk_x dk_y,
\end{equation}
where
\begin{equation} \label{eq:barPhi}
\frac{d^2 \bar{\Phi}}{dz^2} + \left[ \alpha^2 - k^2 + k^2 f(z) \right] \bar{\Phi} = 0
\end{equation}
with $k^2 = k_x^2 + k_y^2$, $k_x$ and $k_y$ the wave numbers in $x$- and $y$-directions. If equation \eqref{eq:fLow} is used, $\bar\Phi$ is expressed in terms of Bessel functions, and if equation \eqref{eq:fNW} is used $\bar\Phi$ is given by hypergeometric functions. 

\citet{Nadol2025} used an asymptotic form of \eqref{eq:fNW} in order to avoid the computationally expensive hypergeometric solutions. In this case the solutions for $\bar\Phi$ are defined by combinations of exponential functions. This asymptotic solution can be used when certain assumptions about $\Delta z$ are made. This solution requires $\Delta z$ small, such that $\omega = (z-z_0)/\Delta z$ becomes either large and positive, or large and negative, depending on the sign of $z-z_0$. If $\Delta z$ is much larger than $z_0$, the absolute value of $\omega$ does not become large for moderate values of $z$. Therefore, $\Delta z$ must be smaller than $z_0$. In this case the asymptotic solution can be obtained under the assumption that equation \eqref{eq:fNW} is approximated by a step function \citep{Nadol2025}.

The three MHS solutions described above are implemented in \texttt{MHSXtraPy}. Potential and linear force-free (LFF) extrapolations can also be calculated by making certain parameter choices: for potential fields one has to choose $\alpha = a = 0$, while $\alpha \neq 0$ in combination with $a=0$ leads to the LFF case, where $a$ refers to either $a_L$ in equation \eqref{eq:fLow} or $a_{N+W}$ in equation \eqref{eq:fNW}.

The extrapolation is based on a photospheric magnetogram as boundary condition at $z=0$ and assumes $\textbf{B} \to 0$ for $z \to \infty$. For handling the boundary condition the code makes use of two assumptions: (1) The photospheric magnetogram used as boundary condition at $z = 0$ is flux-balanced. In cases of a bottom boundary condition for which the magnetogram is not flux-balanced the method suggested by \citet{Seehafer1978} is used to generate a flux-balanced lower boundary condition; (2) The lateral boundary conditions ($x$- and $y$-directions) are assumed to be periodic. Under these assumptions the boundary condition $B_{z, Obs}(x,y,0)$ can be represented by a Fourier series of the form 
\begin{equation}
    B_{z, Obs}(x,y,0) =  \sum_{m= -\infty}^\infty \sum_{n= -\infty}^\infty h_{nm} \exp(ik_nx+ik_my) \label{eq:BC}
\end{equation}
with coefficients $h_{nm} \in \mathbb{C}$, $k_n =  2 \pi n / L_x$ and $k_m =  2 \pi m / L_y$, where $L_x$ and $L_y$ are the normalising length scales in the $x$- and $y$-directions. From this the magnetic field components are obtained from equation \eqref{eq:poltor} \citep[for details see e.g.][]{NadolThesis}.

Generally, using the method described above the magnetic field \textbf{B} can be calculated analytically for all arbitrary $(x, y, z)$. However, the boundary condition will usually be given on a discrete grid. In case of an observational boundary condition $B_{z, Obs}$ this grid is pre-determined and has a finite resolution: If the resolution of this "data grid" is given by $N$ in $x$-direction and $M$ in $y$-direction, the maximal number of Fourier modes in the $x$- and $y$-directions is restricted by $N$ and $M$, respectively. Therefore, the sums in equation \eqref{eq:BC}, and consequently for the magnetic field components, will be truncated. In the code the maximally possible number of Fourier modes is set as default, but a restricted number of modes $N_{F,x} \leq N$ in $x$-direction and $M_{F,y} \leq M$ in $y$-direction can also be selected. \texttt{MHSXtraPy} calculates the coefficients $h_{nm}$ for $1 \leq n \leq N_{F,x}$ and $1 \leq m \leq M_{F,y}$ from the output of a standard numerical Fast Fourier Transform (FFT) calculation. These coefficients can be used together with the solution of equation \eqref{eq:barPhi} to calculate $B_x$, $B_y$ and $B_z$. Restricting $N_{F,x}$ and $M_{F,x}$ as described above could prove useful for future runtime optimisations of the code, as in general, most FFT implementations perform best on certain list sizes as they utilise the algorithm by \citet{Tukey1965} which is optimised for input lengths that are powers of two.


Truncating the Fourier sums does still allow the calculation of the magnetic field and other quantities at arbitrary points $(x, y, z)$. However, to increase the numerical efficiency of the code it only calculates and stores these quantities on a discrete grid of points ("graphics grid"). Interpolation between these grid points is, for example, used to plot magnetic field lines. Although the size of the graphics grid and the size of the data grid (provided by the boundary magnetogram) are independent of each other they are set to have the same resolution in the $x$- and $y$ directions in the current version of the code. The resolution in the $z$-direction of the graphics grid is set to be the maximum height of the computational domain divided by the pixel size of the magnetogram.

\subsubsection{Plasma Pressure and Density}

In the MHS solution plasma pressure and density are given by
\begin{align}
	p(x,y,z) &= p_b(z) - \frac{f(z)}{2\mu_0} B_z^2, \label{eq:p} \\
	\rho(x,y,z) &= \frac{1}{g} \left( - \frac{dp_b(z)}{dz} + \frac{d f(z)}{dz}\frac{B_z^2}{2 \mu_0}  + \frac{f(z)}{\mu_0} \textbf{B} \cdot \nabla B_z \right),\label{eq:rho}
\end{align}
where $p_b$ is an integration "constant" (in $x$ and $y$, therefore allowed to vary with $z$) with $p_b(z) > 0$ for all $z$. Methods to choose $p_b$ include both a priori (as done here) and posteriori options \citep[as for example done in][]{Wiegelmann2015}. Both methods are prone to different difficulties, e.g. negative pressure and density values potentially appearing somewhere in the domain (primarily in the a priori case), or unrealistically large values of pressure and density being generated somewhere in the domain (primarily in the a posteriori case). Therefore, \texttt{MHSXtraPy} is primarily used to obtain the deviations in pressure and density from the background pressure $p_b$ and background density $\rho_b = - \frac{dp_b(z)}{gdz}$. These deviations from hydrostatic balance are explicitly given by 
\begin{align}
	\Delta p &= - \frac{f(z)}{2\mu_0} B_z^2, \label{eq:dp}\\
	\Delta \rho &=  \frac{1}{g} \left( \frac{d f(z) / dz}{2 \mu_0} B_z^2 + \frac{f(z)}{\mu_0} \textbf{B} \cdot \nabla B_z \right). \label{eq:dd}
\end{align}
Additionally, \texttt{MHSXtraPy} provides two different background atmosphere models, but when using these one needs to ensure that the full pressure and density stay positive and do not become unrealistically large. 

The first one is based on a background temperature profile represented by a hyperbolic tangent connecting a photosphere of nominal value $5600$ K and a corona of nominal value $2.0 \times 10^{6}$ K. The second one uses linear interpolation between given temperatures at given heights. In both cases, pressure and density are obtained using the ideal gas law together with the Boltzmann constant $k_B = 1.380649 \times 10^{-23}$ J K$^{-1}$, mean molecular weight $\bar\mu$ and solar gravitational acceleration $g$.

\begin{figure} 
 \centering
     \includegraphics[width=0.23\textwidth]{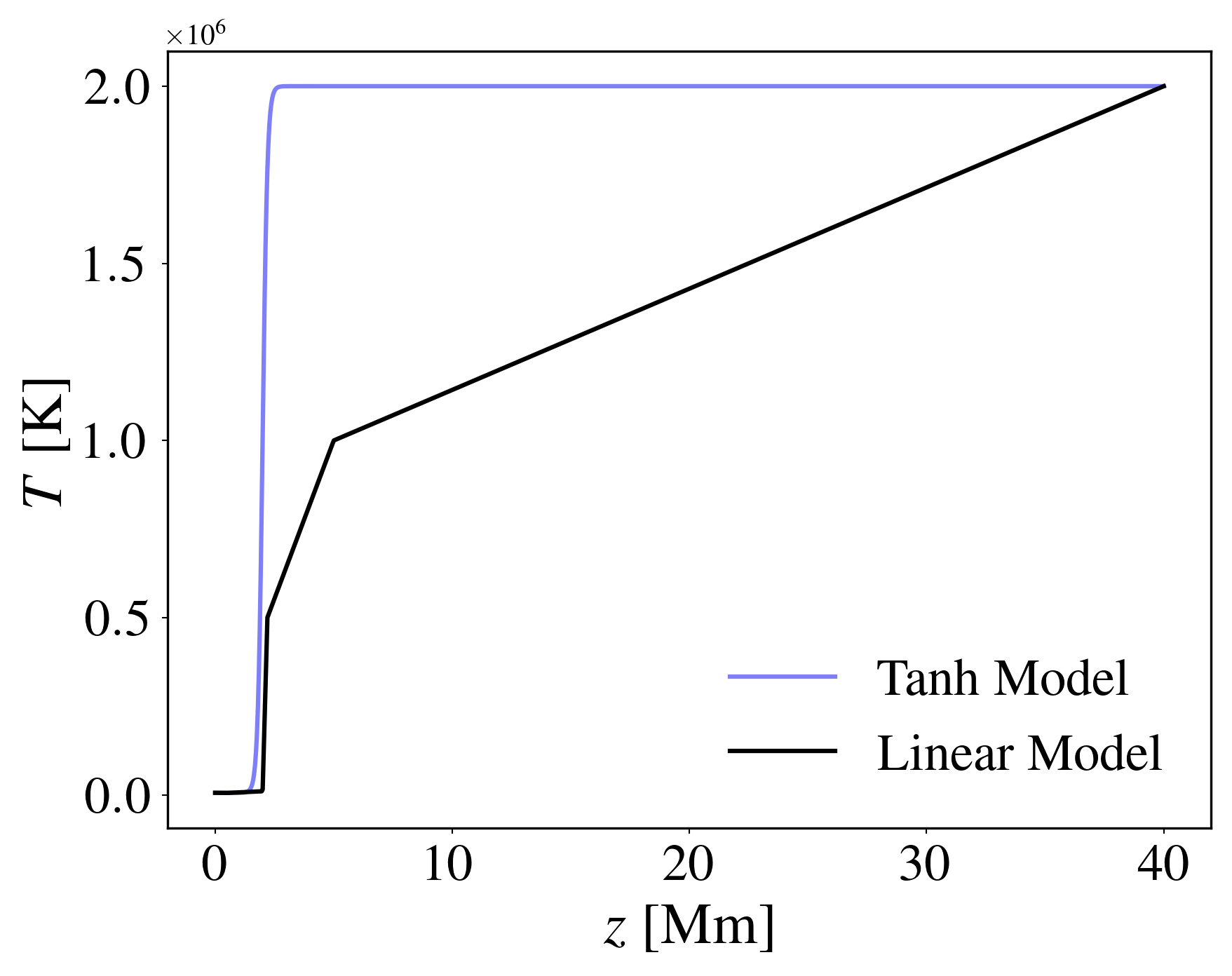}
     \includegraphics[width=0.23\textwidth]{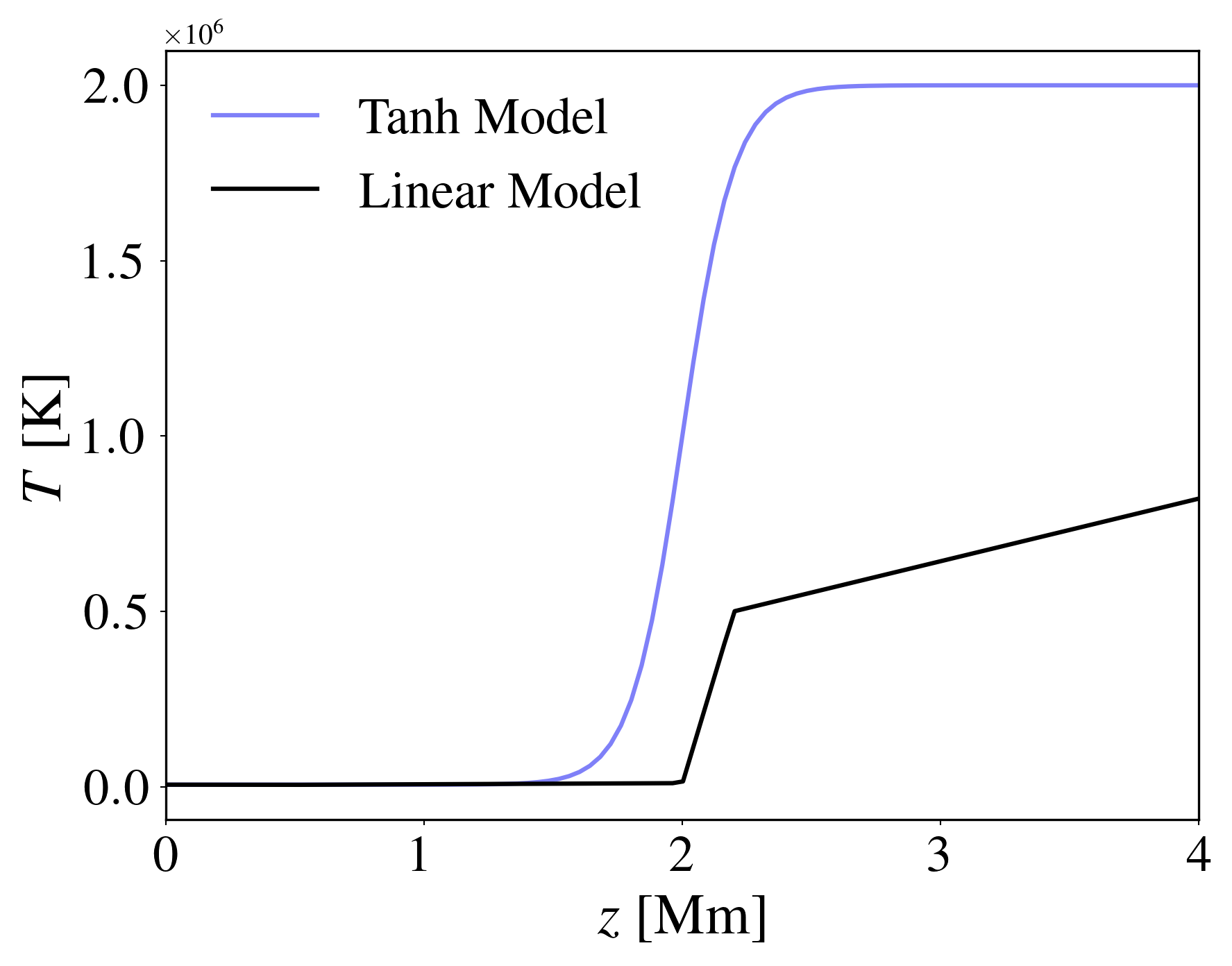}
     \includegraphics[width=0.23\textwidth]{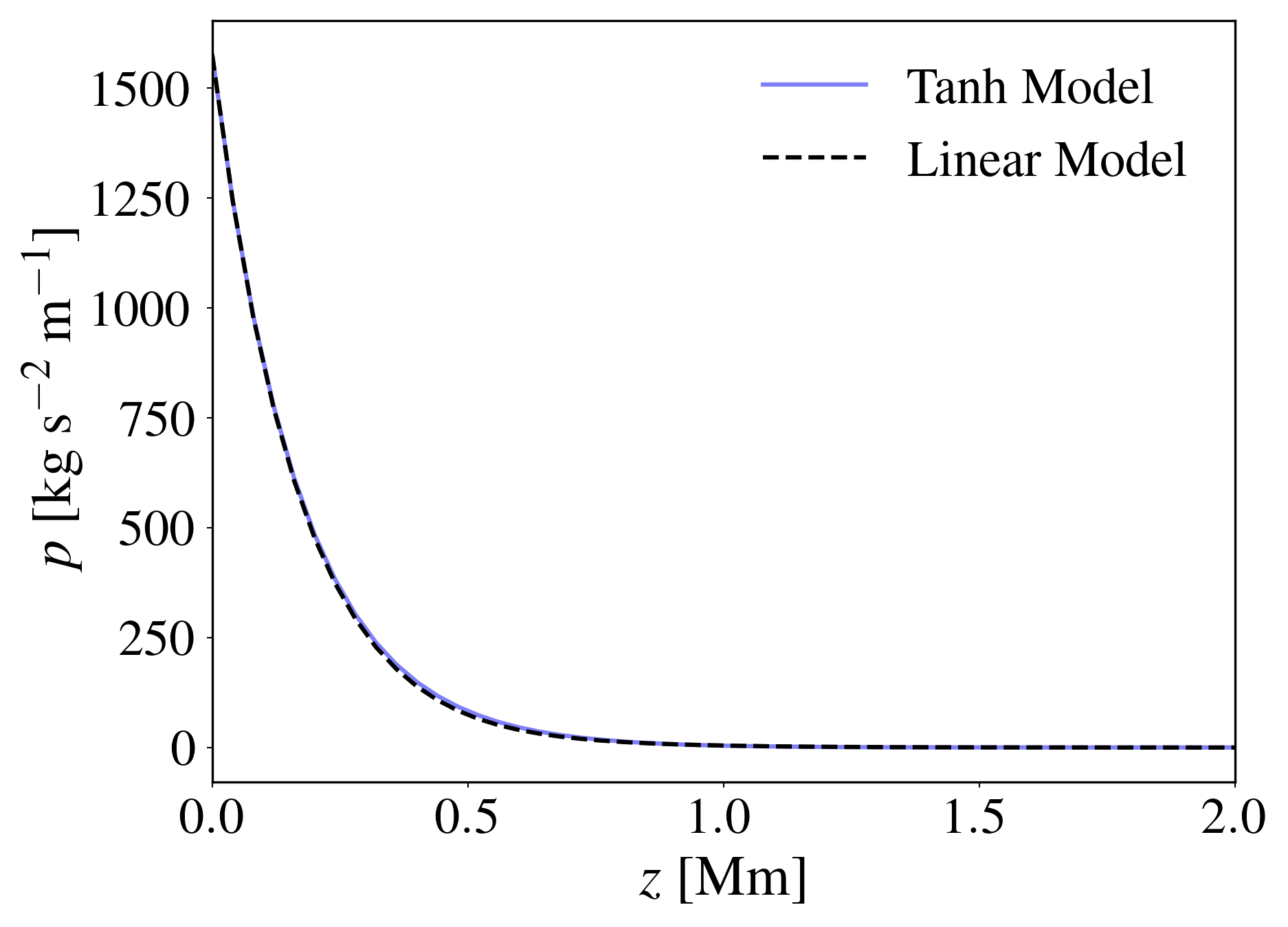}
     \includegraphics[width=0.23\textwidth]{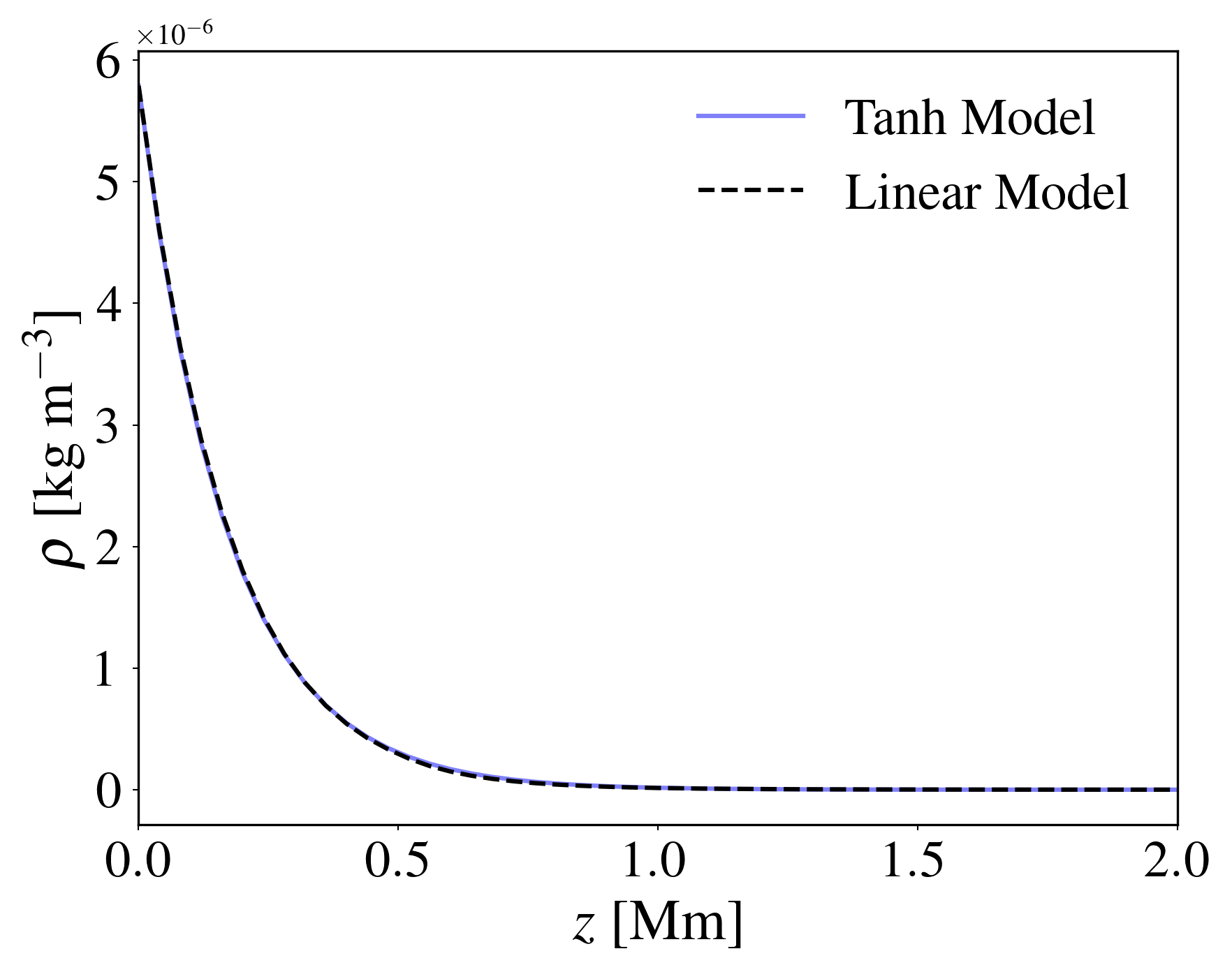}
     \caption{Examples for the two background atmospheres implemented in \texttt{MHSXtraPy}. The two figures in the top row show the temperature curves both for the full extrapolation height $40$ Mm (on the left) and for a region around the transition region (up to $4$ Mm above the photosphere). The hyperbolic tangential temperature profile (blue) is calculated with photospheric temperature $5600$ K and coronal temperature $2$x$10^6$ K. The linear temperature profile (black) is interpolated between $5600$ K, $5200$ K, $10^4$ K, $5x10^5$ K, $10^6$ K and $2$x$10^6$ K at heights $0$ Mm, $0.5$ Mm, $2.0$ Mm, $2.2$ Mm, $5.0$ Mm and $40.0$ Mm above the photosphere. The two figures in the bottom row show the resulting pressure and density curves, respectively, for the two temperature profiles.} 
     \label{fig:backatm-aia}
 \end{figure}

Examples for both background atmosphere models can be seen in Figure \ref{fig:backatm-aia}. The first model, which uses a hyperbolic tangential function for the temperature profile, is shown in blue. While this temperature profile is not fully realistic, it allows for modelling a sharp increase in temperature at the height of the transition region, here at $z=2$. The linear interpolation model, here shown in black, is an empirical choice motivated by observed values of temperature in the solar atmosphere and letting the user introduce as many given steps as desired. It was included to provide an alternative to the purely analytical model.

Generally, any background model requires $dp_b/dz$ strictly negative in the whole domain; this excludes for example the model by \citep{Vernazza1981}.

\subsection{Example} \label{sec:Example}

\begin{figure} 
 \centering
     \includegraphics[width=0.49\textwidth]{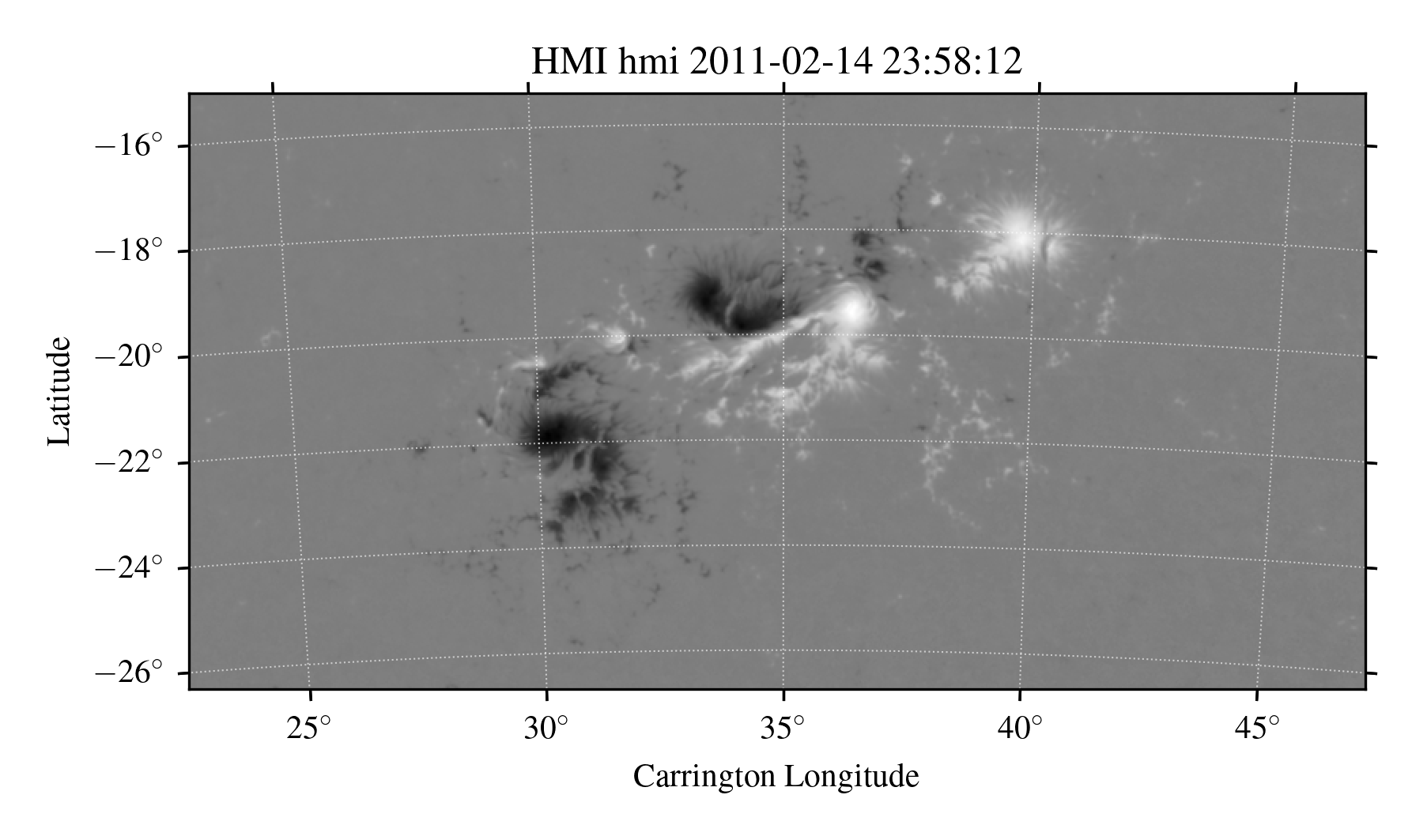}
     \includegraphics[width=0.49\textwidth]{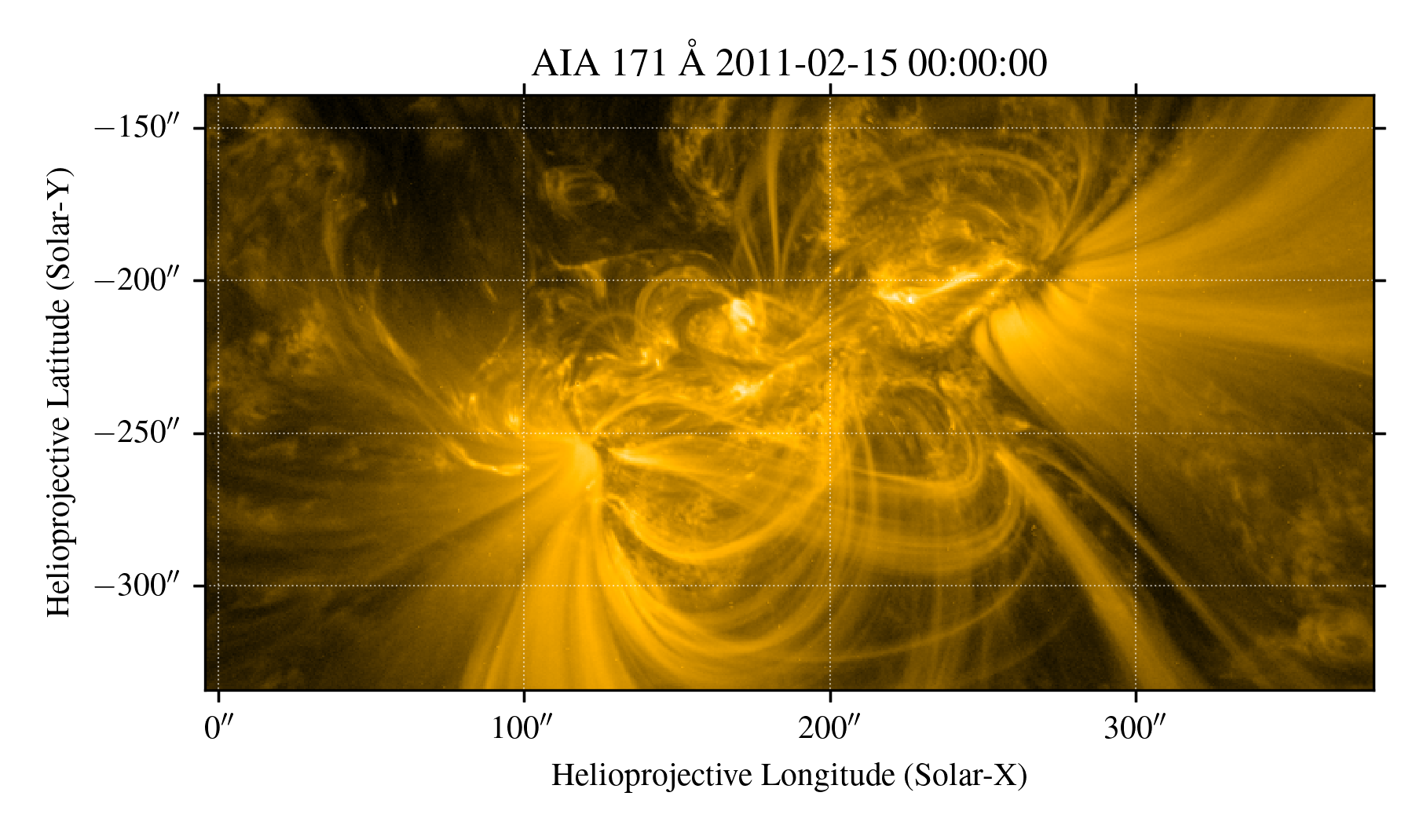}
     \caption{Magnetic and coronal observations of the investigated active region on the Sun. Top panel: Radial SHARP magnetogram as observed by SDO/HMI on 2011-02-14 at 23:58:12 TAI. Bottom panel: Extreme ultraviolet image from the Atmospheric Imaging Assembly (AIA) at 171 \AA, taken shortly after on 2011-02-15 at 00:00:00 TAI, displaying the coronal loops associated with the active region.} 
     \label{fig:sharp-aia}
 \end{figure}

In this section we demonstrate the functionality of the code by presenting an example application using an SDO/HMI observation of active region NOAA 11158 (SHARP 377) as boundary condition. This region emerged on 2011-02-12 and was observed over 5 days by SDO/HMI. The snapshot used as boundary condition here was observed at midnight TAI\footnote{International Atomic Time, see e.g. \url{https://gssc.esa.int/navipedia/index.php/Atomic_Time}.} on 2011-02-14; both the HMI magnetogram as well as a matched AIA 171 observation are shown in figure \ref{fig:sharp-aia}.


\subsubsection{Data preparation}

SDO/HMI observes full disk magnetograms, such that the obtained image of the line-of-sight magnetic field component needs to be cut to an area around the (active) region, that one aims to investigate. As \texttt{MHSXtraPy} uses Cartesian geometry, it is important that the region of interest is close to the disk centre and the cutout small enough that neglecting the curvature of the solar surface is appropriate and the utilisation of such geometry justified. Additionally, while cutouts of smaller size are favourable for shorter runtimes, they may also lead to interesting features and information getting lost. This trade-off between increased numerical efficiency and minimising the effects of the boundaries needs to be considered when using observational data as input. 

Specifically, it is advantageous if the active region under investigation is reasonably well isolated. Users of the code could consider to modify the boundary condition at $z=0$ provided by an observed magnetogram by adding a zone in the $x$- and $y$- directions, a few pixels in width, with $B_z = 0$. While this increases the grid size the method is fast enough to allow for this adjustment. However, this essentially forces the field lines to be horizontal above the added regions.

For the example shown here we have used a preprocessed SHARP region that fulfils the above criteria. The used cutout extends approximately 271 Mm in $x$-direction (longitudinal) and 137 Mm in $y$-direction (latitudinal). Its magnetic field strength ranges from $-2565.05$ to $2617.67$ Gauss. 

\subsubsection{Solution selection and parameters}

As mentioned above the code allows for the application of three different solutions: (1) the solution in which the transition from non-force-free to force-free is controlled by equation \eqref{eq:fLow} \citep{Low1991}; (2) the solution in which this transition is controlled by equation \eqref{eq:fNW} \citep{Neukirch2019}; and (3) the solution in which this transition is controlled by the step function approximation of equation \eqref{eq:fNW} \citep{Nadol2025}.

For the example here, we have chosen solution (3). In this case the parameters $\alpha$, $a_{N+W}$, $z_0$ and $\Delta z$ need to be specified.\footnote{The approximation of equation \eqref{eq:fNW} by a step function eliminates $\Delta z$ from the expression for $f(z)$ in the calculation of \textbf{B}. Nevertheless, $\Delta z$ is still needed for the calculation of \textbf{j}, $p$ and $\rho$. Here $f(z)$ is directly used and therefore, not asymptotically approximated.} Theoretically, the parameter $\alpha$ can be chosen freely and $a_{N+W}$ can be chosen depending on the value of $\alpha$ as it is bounded by
\begin{equation*}
	a_{N+W, max} = \frac{k_{min}^2-\alpha^2}{(1+b)k_{min}^2},
\end{equation*}
where $k_{min}^2 = \min \left\{ k^2 = k_n^2 + k_m^2 \hspace{0.5mm} | \hspace{0.5mm} \forall \hspace{0.5mm} n, m \right\}$. While $\alpha$ is responsible for how much the resulting field lines twist, $a_{N+W}$ specifies the amplitude of the current perpendicular to the magnetic field on the photosphere. As both these effects are frequently observed on the Sun, we want to choose $\alpha$ and $a_{N+W}$ in practice in a way, that the effect of both are balanced and both resulting phenomena are included in the model to a meaningful extent.

If one wanted to prioritise either of the effects, the respective value must be chosen larger (and the other value smaller). While this is possible, it might be advantageous to use a non-linear force-free model in the case of strongly twisted magnetic field, but as a consequence no perpendicular currents are included in the model. Ideally, a non-linear MHS solution would be used, but this requires significantly larger computational resources, and usually longer runtimes. 

General methods on how to determine the parameter $\alpha$ have been proposed for example by \citet{Leka1999} or \citet{Hagino2004}. The latter one has been implemented in the code. These methods can be applied when measurements of all three vector components are available on the photosphere. If only line-of-sight observations are available one must use other approaches to finding a good value of $\alpha$: e.g. one can match the magnetic field geometry to other EUV observations to find a value of $\alpha$ for which the extrapolated field lines match best with visible structures \citep[see e.g.][]{Wiegelmann2023}. Generally, applying the MHS model in cases where the "optimal" value of $\alpha$ leads to very small values of $a$, makes the use of the model redundant compared to linear force-free models.

In this example, we choose $\alpha = 0.01$ for illustrative purposes as this allows for relatively large values of $a_{N+W}$ as $a_{N+W, max} = 0.4069$ (for $b=1$). We have chosen $a_{N+W}=0.4$. The other two parameters $z_0$ and $\Delta z$ are chosen such that they agree with commonly observed values. The centre of the region over which the transition from non-force-free to force-free occurs is set to $z_0 = 2$ Mm, approximately the upper chromosphere, and the width over which it  takes place is set to $\Delta z = 0.2$ Mm, approximately the width of the transition region.

\subsubsection{Results}

\begin{figure} 
 \centering
     \includegraphics[width=0.49\textwidth]{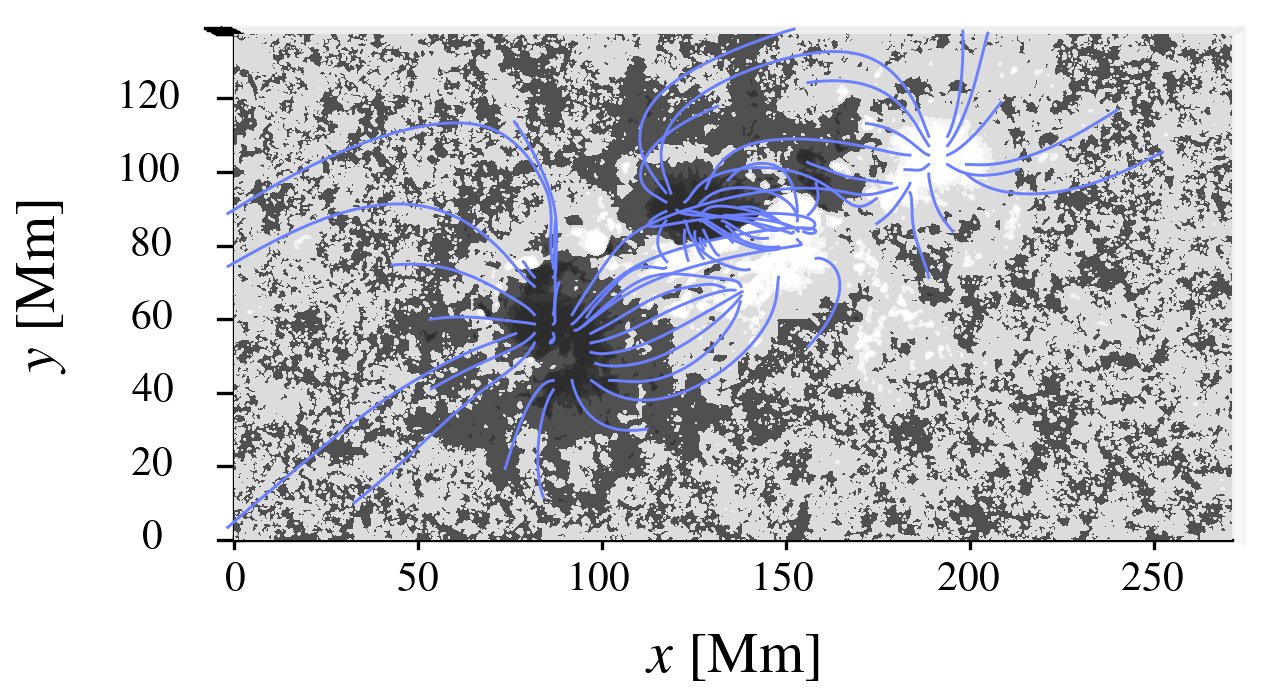}
     \includegraphics[width=0.49\textwidth]{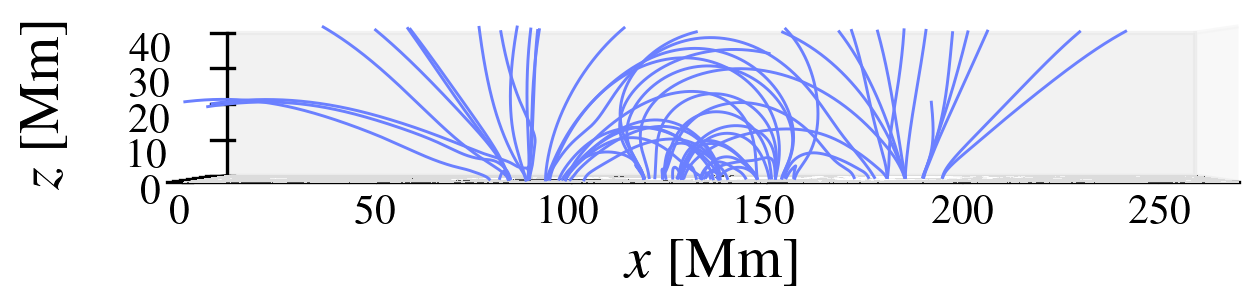}
     \includegraphics[width=0.49\textwidth]{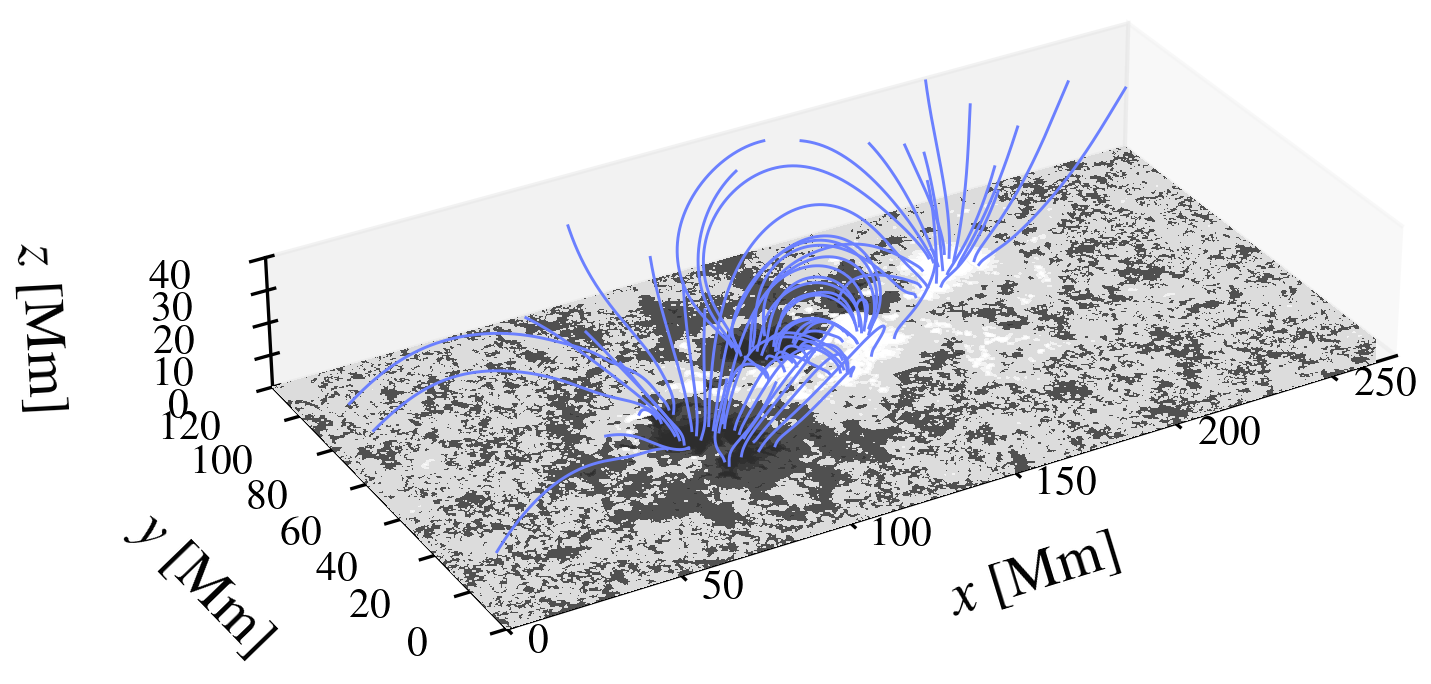}
     \caption{Field lines visualised by \texttt{MHSXtraPy} resulting from the magnetic field extrapolation above active region NOAA 11158.}
     \label{fig:magnetogram-3D}
 \end{figure}

\begin{figure} 
 \centering
     \includegraphics[width=0.49\textwidth]{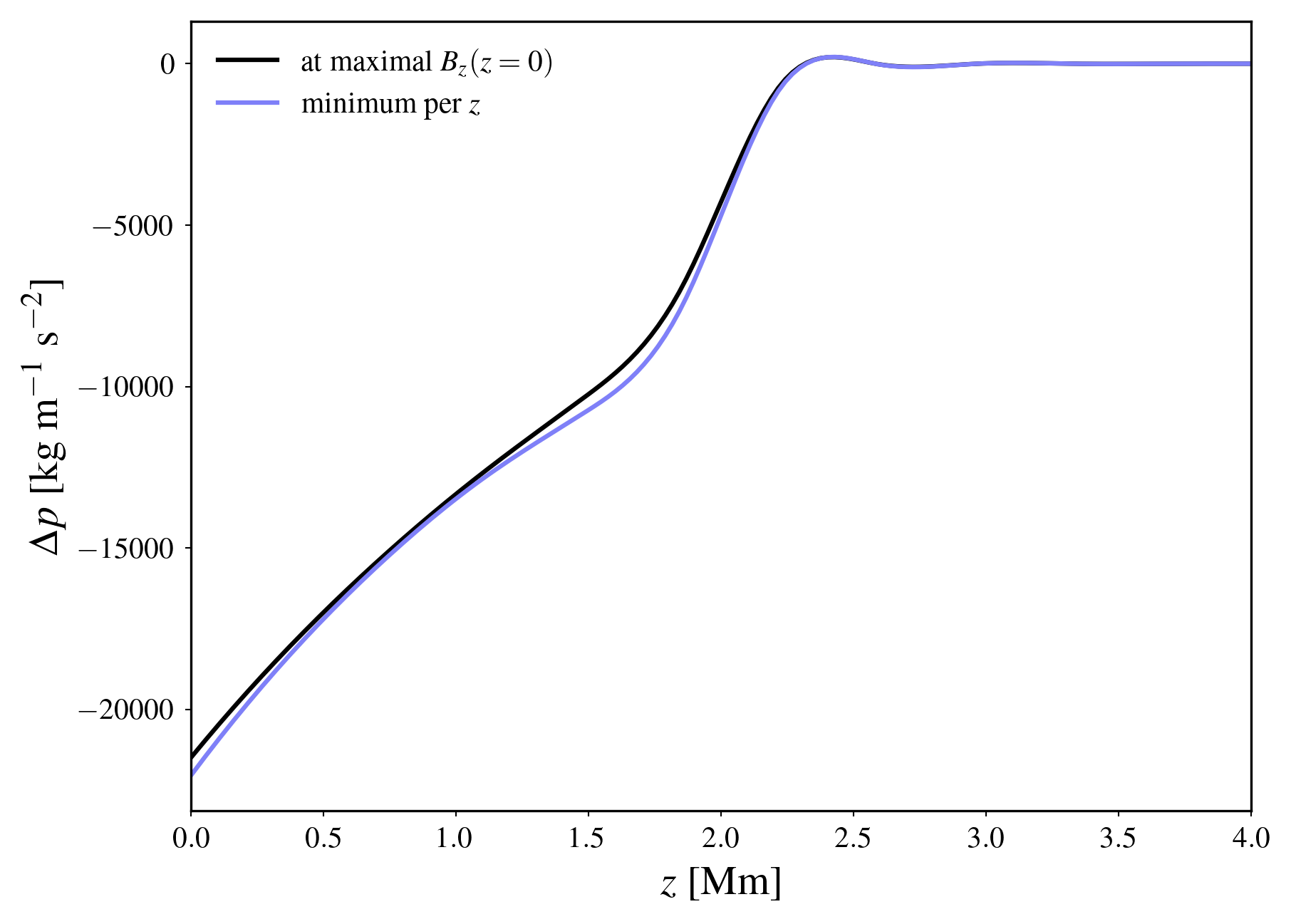}
     \includegraphics[width=0.49\textwidth]{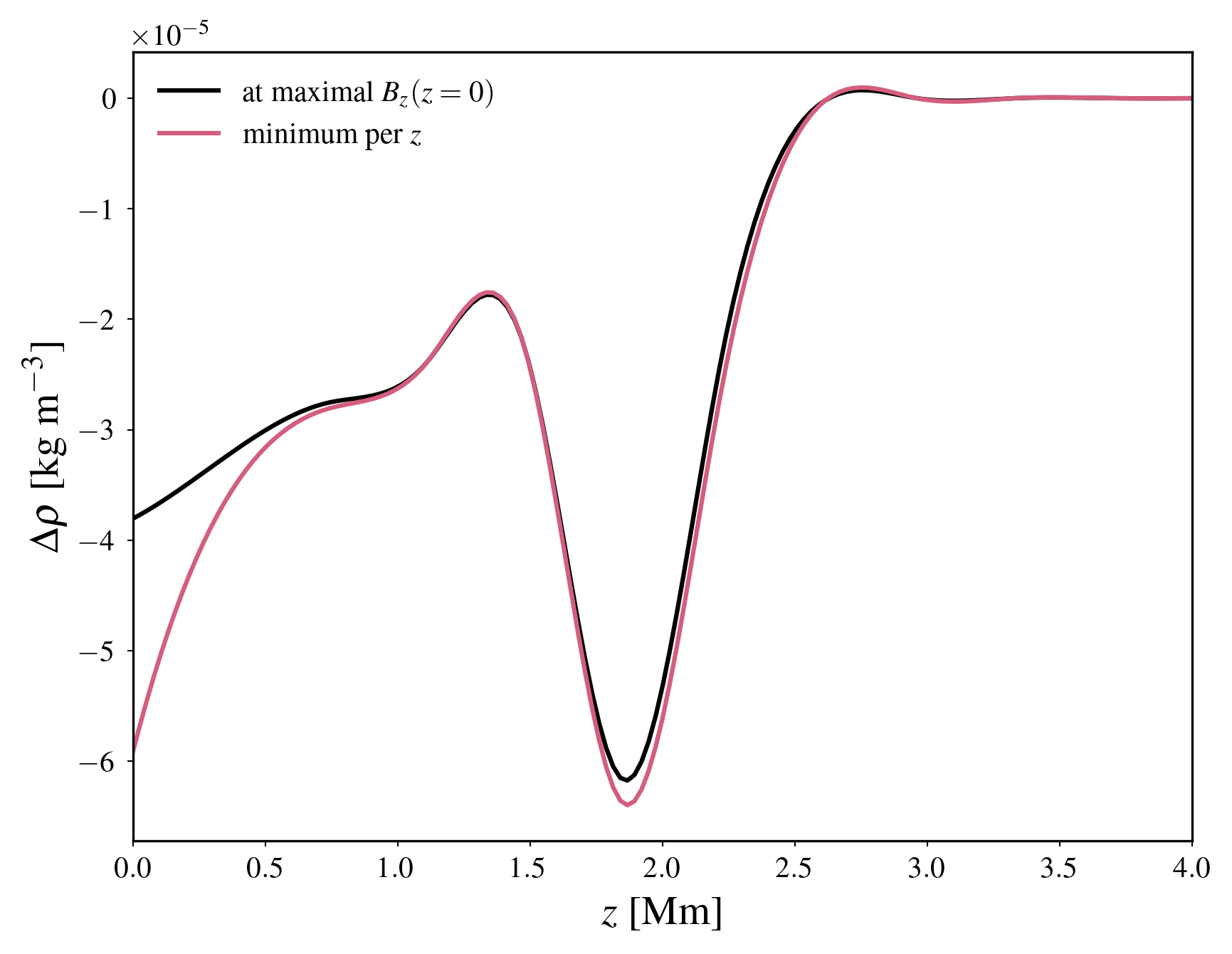}
     \caption{Variations in plasma pressure and density with height $z$ above the photosphere, between $z=0$ Mm and $z=4$ Mm. The change from non-force-free to force-free occurs between approximately $z=1.8$ and $z=2.2$ Mm, which is particularly obvious in the density variation.} 
     \label{fig:plasmaparam-z}
 \end{figure}

\begin{figure} 
 \centering
     \includegraphics[width=0.49\textwidth]{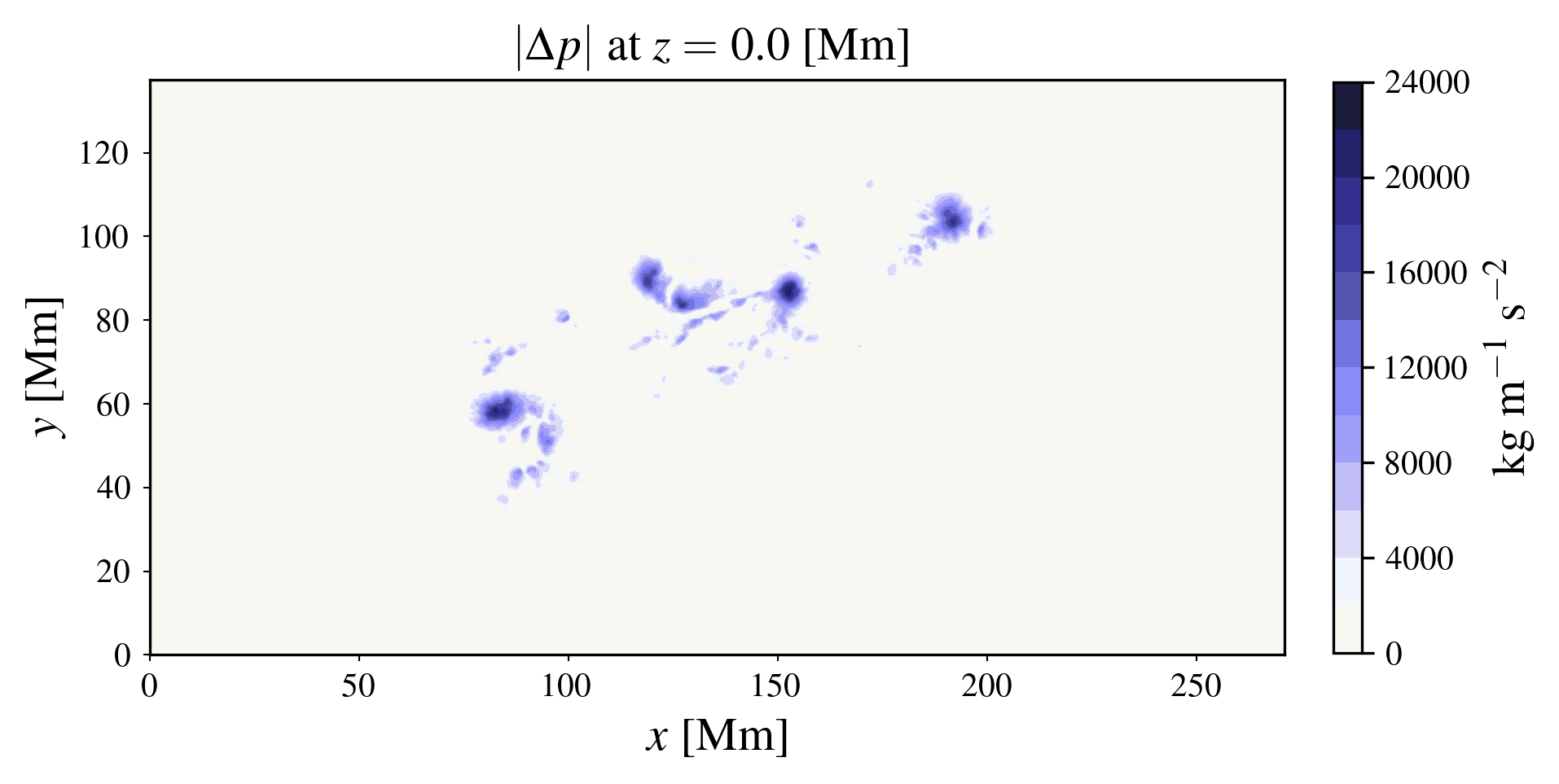}
     \includegraphics[width=0.49\textwidth]{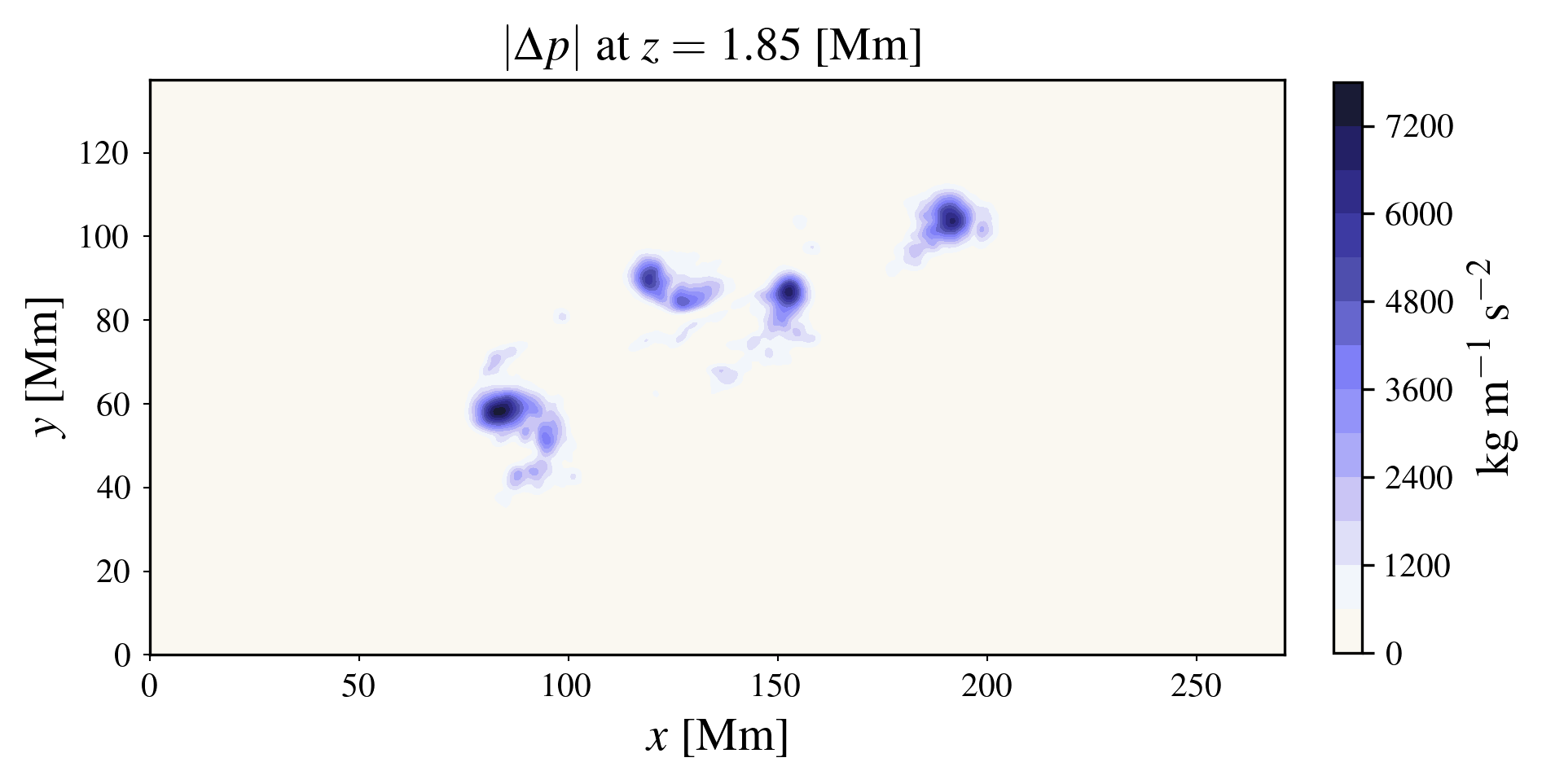}
     \includegraphics[width=0.49\textwidth]{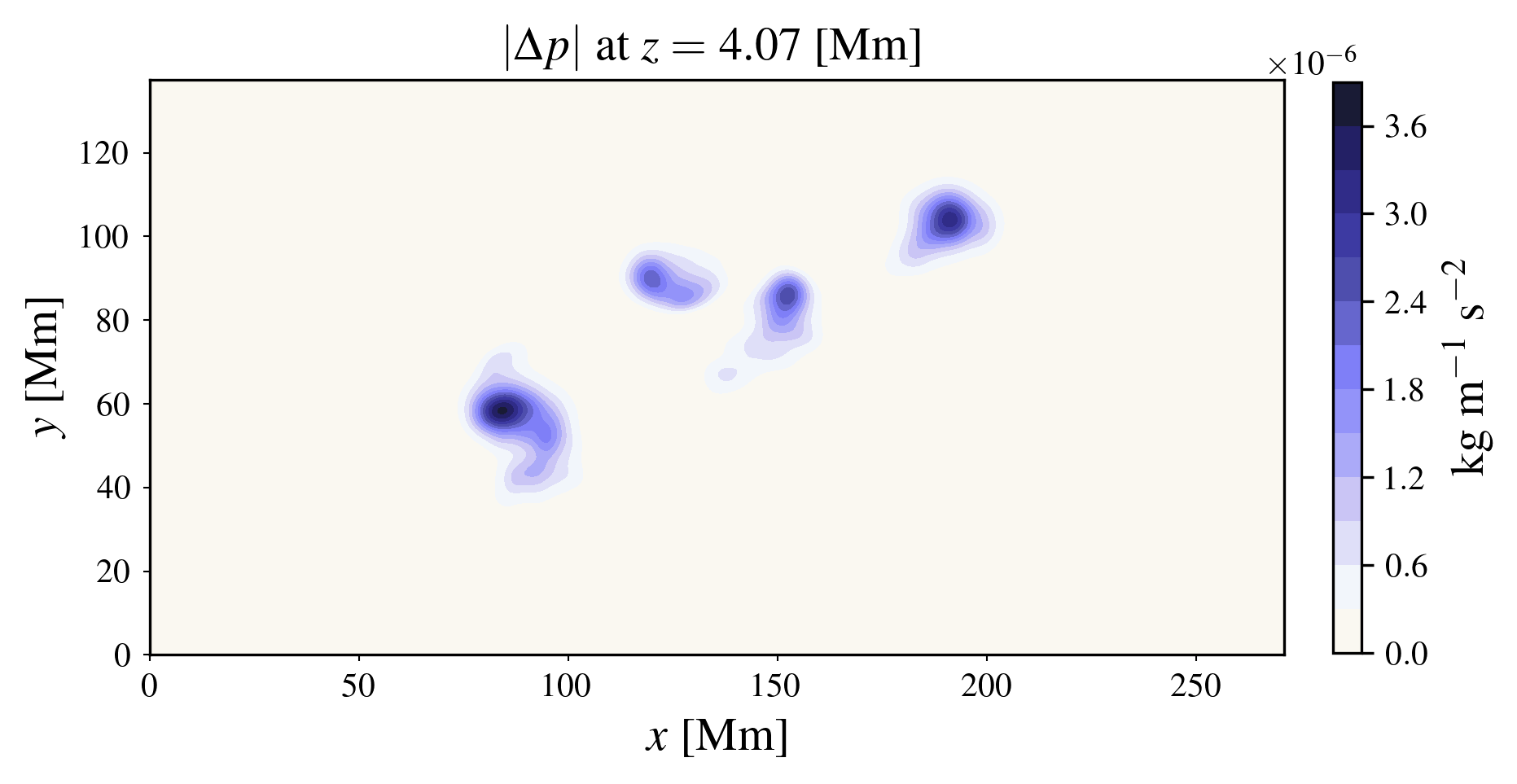}
     \caption{Variation in plasma pressure at different heights across the whole horizontal domain. The observed heights are $z =0$ (the photosphere), $z=1.85$ (approximately the height of the transition region), and $z=4.07$ (in the lower corona). In the lower corona observations and values of $\beta_p$ suggest that an almost force-free state should be reached. The model represents this well as seen in the scale being multiplied by $10^{-6}$ in the bottom panel.} 
     \label{fig:plasmapress}
 \end{figure}

\begin{figure} 
 \centering
     \includegraphics[width=0.49\textwidth]{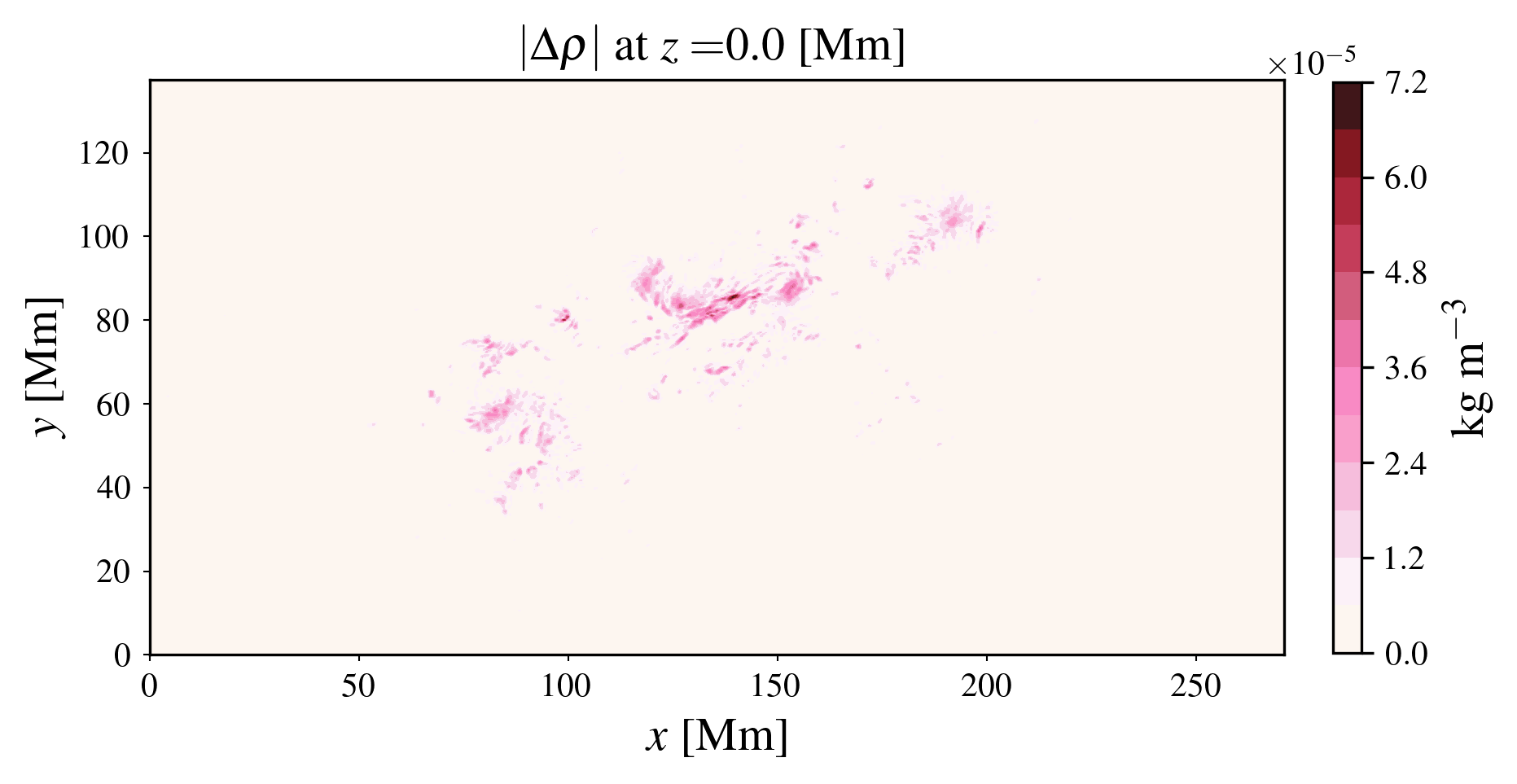}
     \includegraphics[width=0.49\textwidth]{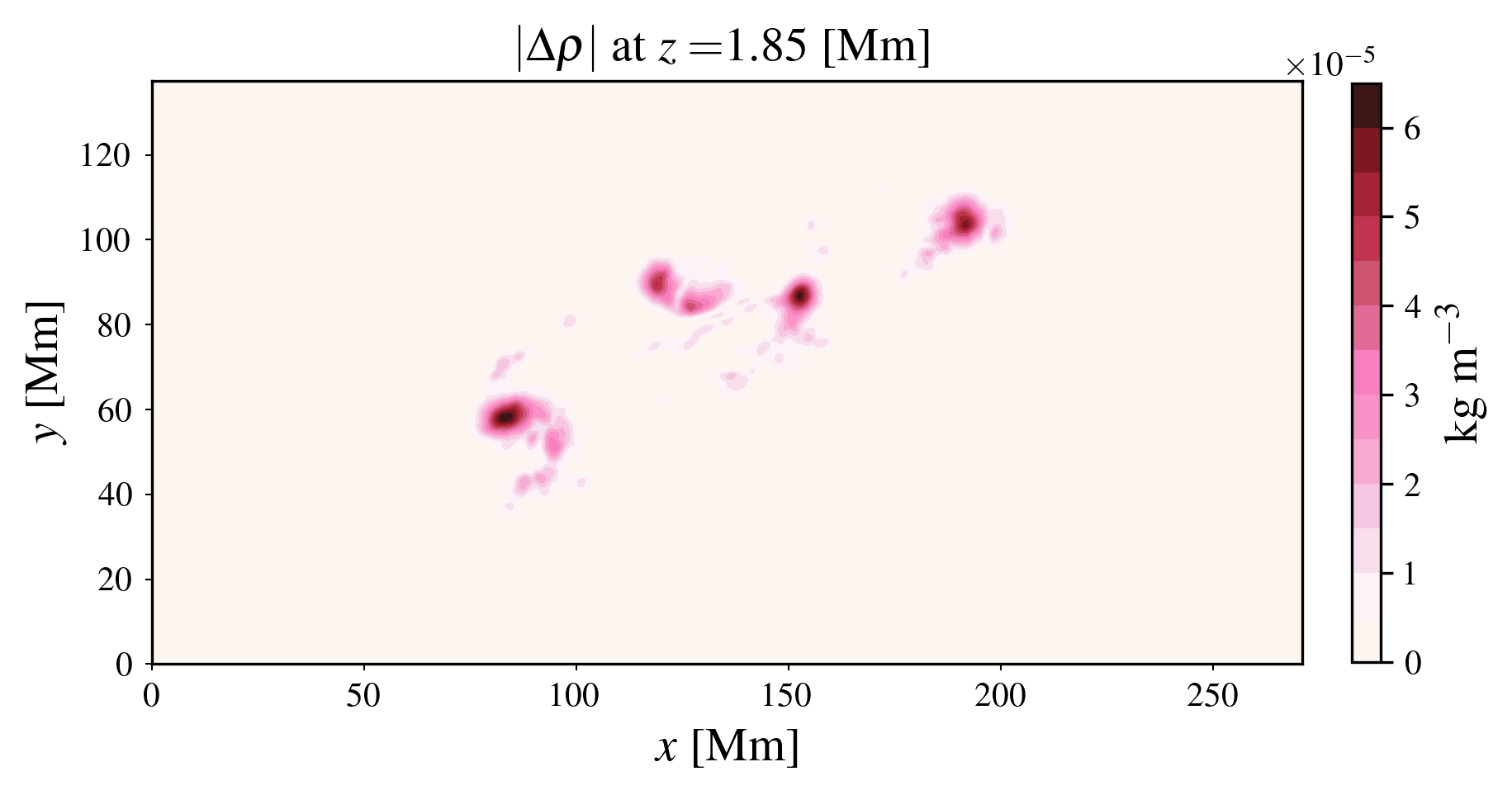}
     \includegraphics[width=0.49\textwidth]{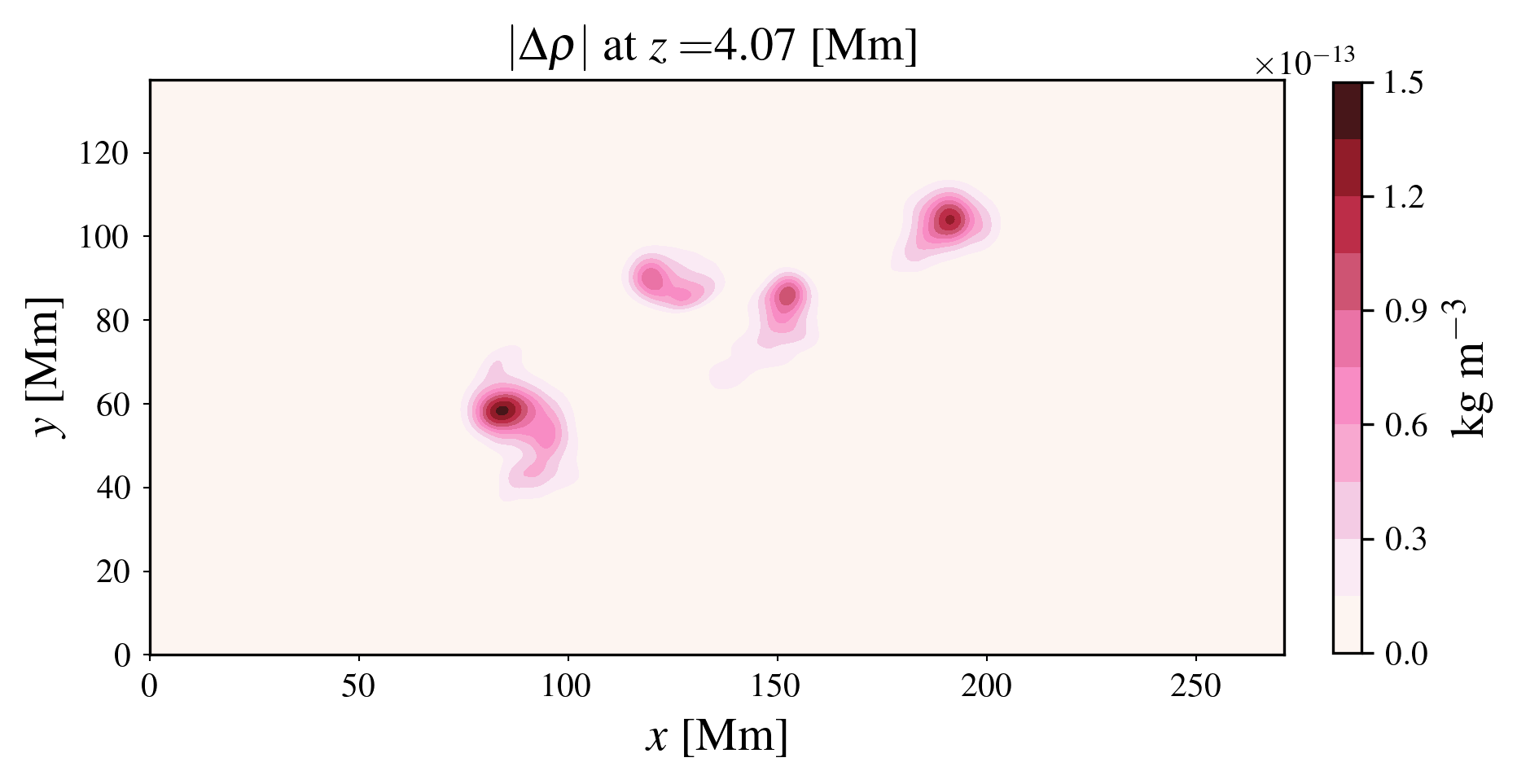}
     \caption{Variation in plasma density at different heights across the whole horizontal domain. The observed heights are $z =0$ (the photosphere), $z=1.85$ (approximately the height of the transition region), and $z=4.07$ (in the lower corona). In the lower corona observations and values of $\beta_p$ suggest that an almost force-free state should be reached.}
     \label{fig:plasmaden}
 \end{figure}

The results of our extrapolation calculation from the chosen boundary condition are shown in Figures \ref{fig:magnetogram-3D}, \ref{fig:plasmaparam-z}, \ref{fig:plasmapress} and \ref{fig:plasmaden}.

Figure \ref{fig:magnetogram-3D} shows extrapolated field lines above the region of interest from above, the side and from with an angular view onto the photosphere to emphasise the 3D nature of the model. 

The model calculates field lines leaving the box\footnote{This refers to field lines seen as open in our computational domain. For this example the extrapolation domain is defined by $(x,y,z) \in [0, 271.13] \times [0, 137.39] \times [0, 40]$ Mm with resolution $744 \times 377 \times 109$.} above the regions of stronger absolute magnetic field as well as field lines closed inside the computational box between areas of opposite polarity. While the standard extrapolation height in the code is kept at $20$ Mm such that the photosphere, chromosphere and the lower corona are included, we have chosen an upper boundary of $z_{max} = 40$ Mm in this example to showcase the possibilities the code has to offer. A higher upper boundary can lead to display field lines closed inside the box that extend higher into the atmosphere connecting negative and positive polarities further apart than just in the centre of the plane as seen in the middle panel of figure \ref{fig:magnetogram-3D}. Larger extrapolation heights might provide valuable insight into the structure of coronal loops, flares and CMEs. However, an even higher upper boundary suggests an intent to investigate the magnetic field further away from the Sun rather than closer to its surface rendering a NLFF model probably more suitable, as the effects of the non-force-free regions around $z_0 = 2$ Mm become more and more negligible as the extrapolation height $z_{max}$ becomes larger.  

 Figure \ref{fig:plasmaparam-z} shows the variation in plasma pressure and density resulting from the model at $(x, y)$ where $B_z$ is maximal on the photosphere as well as the minimal variation at each $z$. The variations of both quantities with height $z$ are shown between the photosphere and $4$ Mm above it. Therefore, the centre of the plot coincides with the centre of the region over which the transition from non-force-free to force-free takes place. It is obvious that above $z= 2$ Mm the model approaches a force-free state as both pressure and density variation quickly tend towards zero here. 
 
 Figures \ref{fig:plasmapress} and \ref{fig:plasmaden} similarly show the pressure and density deviation from hydrostatic balance. Each panel in the figures corresponds to a different height above the photosphere (top to bottom: photosphere, transition region, lower corona) and shows how the plasma and pressure varies in the horizontal direction. The largest deviations appear around the positions of positive and negative polarities on the bottom boundary at $z=0$. Consistent with figure \ref{fig:plasmaparam-z} the pressure drops off with height and the density has a minimum around $z=2$ Mm. 

\section{Performance}

The relevant quantities for the model (the magnetic field \textbf{B}, the current density \textbf{j}, as well as plasma pressure $p$ and plasma density $\rho$) are multi-dimensional arrays. Such arrays are particularly well handled by the NumPy library \citep{numpy2020}, which features a rich interface for this data structure, making Python the natural choice for the development of \texttt{MHSXtraPy}. NumPy provides the class \texttt{ndarray}, which supports a large variety of operations that are essential for the time-efficient calculation of our analytical solution. Despite Python being a high-level, scripting language, \texttt{ndarray} operations are implemented natively and provide good performance due to easy vectorisation.

Next to NumPy, Python contains other scientific libraries advantageous for our purpose, including SciPy \citep{scipy2001, scipy2020}, Matplotlib \citep{hunter2007}, SunPy \citep{Sunpy2020, sunpy2023} and Astropy \citep{astropy2013, astropy2018, astropy2022}. Additionally, Numba \citep{numba2015}, an open source just-in-time (JIT) compiler that translates a subset of Python and NumPy code into fast machine code, is used. By using the industry-standard LLVM compiler library, Numba enables compiled numerical algorithms in Python to approach the speeds of C or Fortran. This can be achieved by the user through a simple application of decorators rendering Numba an essential tool for runtime optimisation.

While vectorisation and parallelisation have been applied partly and have already lead to improvements in numerical efficiency, the potential for computational optimisation of the code is not yet exhausted. Especially the utilisation of parallelisation could be extended significantly: \texttt{MHSXtraPy} has so far only been executed using the Central Processing Unit (CPU) on a standard laptop computer as the JIT compiler used for the parallelisation of the code only optimises CPU performance, but does not utilise the Graphics Processing Unit (GPU). Therefore, further improvements could be expected through GPU use. This in combination with an additional increase in storage efficiency might allow for even more time-efficient computation with larger data sets.

\subsection{Runtime test}

To give an idea of the scale of domain that is practical to use with \texttt{MHSXtraPy} we time the execution of the calculation of the magnetic field using the IPython magic shell \texttt{\%timeit}. The following test has been carried out on a MacBook Air (2020) M1 processor with 16 GB RAM.

\begin{figure} 
    \centering
    \includegraphics[width=0.49\textwidth]{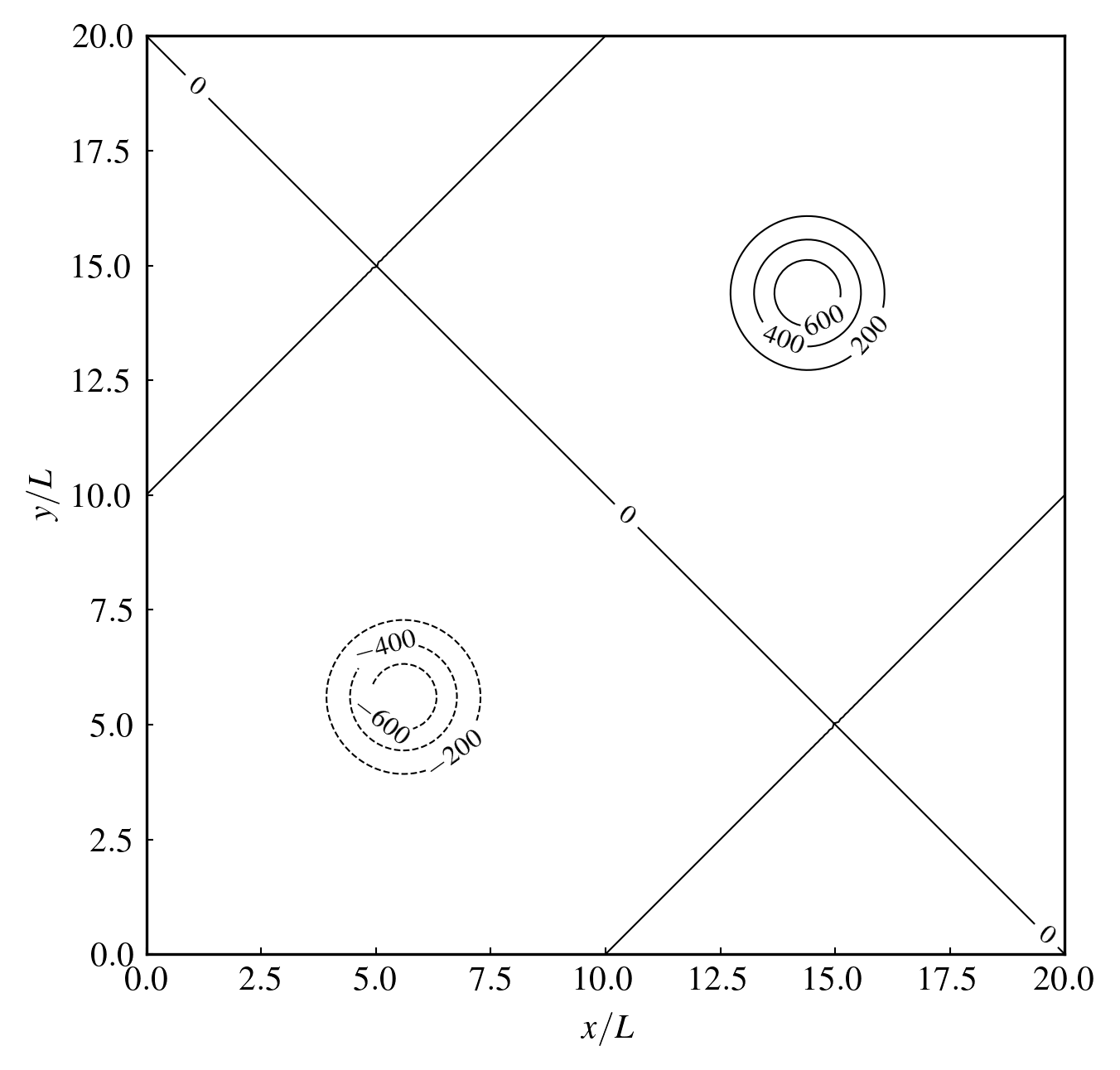}
    \caption{Artificial photospheric magnetic field $B_z(x,y,0)$ in Gauss, as described in equation \eqref{eq:VonMises} for a normalising magnetic field strength $B_0 = 500$ G. Solid lines for positive values (sources), dashed lines for negative values (sinks).}
    \label{VonMises}
\end{figure}

We used \texttt{\%timeit} to estimate the runtime of the function \texttt{b3d}, which can be found in the \texttt{MHSXtraPy} module of the same name. This function calculates all magnetic vector field components, $B_x$, $B_y$, $B_z$, as well as the partial derivatives of $B_z$, which are necessary to calculate the plasma pressure \eqref{eq:rho}. This test was carried out using a simple, artificial initial condition of size $n^2$ given by
 \begin{align} \label{eq:VonMises}
	\frac{B_z(x,y,0)}{B_0} &= \frac{\exp(\lambda \cos(\tilde{x}-\mu))}{2 \pi \text{I}_0(\lambda)} \frac{\exp(\lambda \cos(\tilde{y}-\mu))}{2 \pi \text{I}_0(\lambda)} \nonumber \\ &- \frac{\exp(\lambda \cos(\tilde{x}+\mu))}{2 \pi \text{I}_0(\lambda)} \frac{\exp(\lambda \cos(\tilde{y}+\mu))}{2 \pi \text{I}_0(\lambda)}
 \end{align}
 where $\tilde{x} = \pi ( x / 10 - 1)$ and $\tilde{y} = \pi ( y/ 10-1)$, $\mu = 1.2 / \pi + 1$ and $\lambda = 10$, for $x, y \in [0, 20]$, and $\text{I}_0$ is a modified Bessel functions of the first kind. In Figure \ref{VonMises} we show the contour lines of this boundary condition calculated for a normalising magnetic field strength $B_0 = 500$ G.
 
The discretisation $n$ of the square, two-dimensional bottom boundary is increased during the test, such that the time complexity of the algorithm can be estimated. We have used $n$ from $50$ to $500$ with a step size of $50$. At the same time the vertical resolution $n_z = 200$ is kept fixed, such that the overall number of volume cells is $200 n^2$ in each step. The runtime is averaged over 5 repeats of 5 loops each, resulting in overall 25 executions of \texttt{b3d} per configuration.
 
\begin{figure}
\centering
	\includegraphics[width=0.49\textwidth]{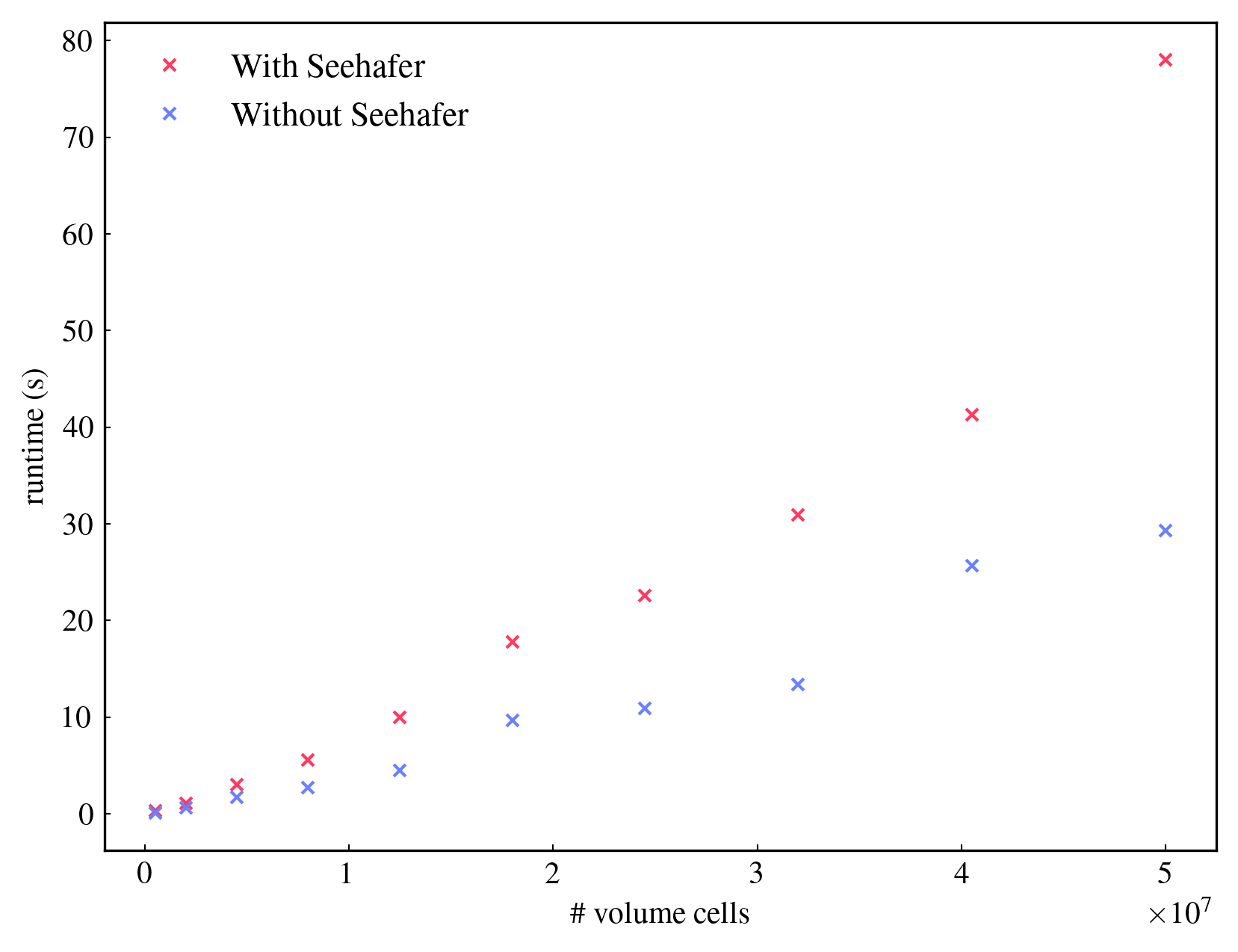}
	\includegraphics[width=0.49\textwidth]{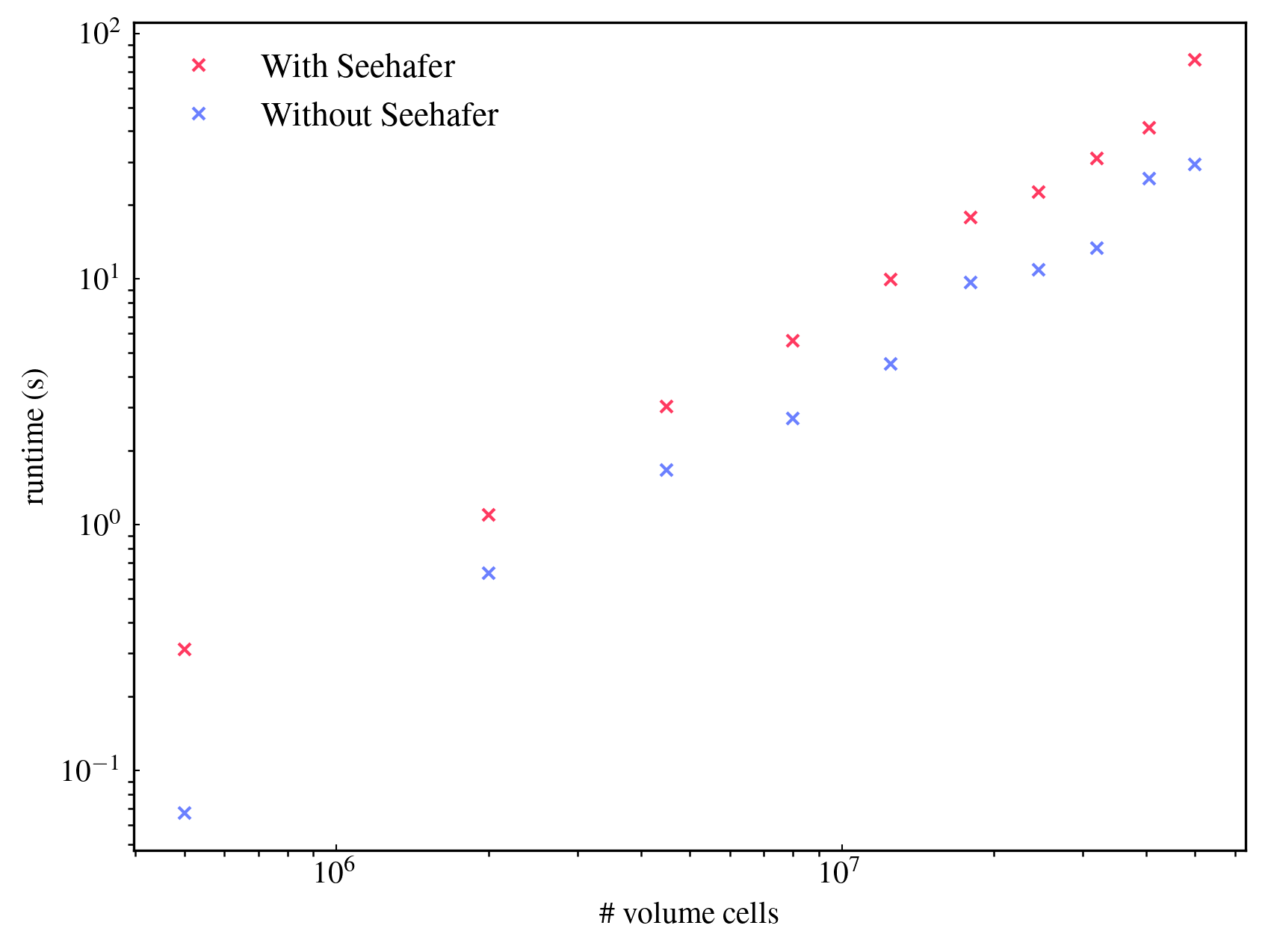}
	\caption{Cumulative runtime of the calculation of $\textbf{B}$ and $\nabla B_z$ using the solution by \citet{Nadol2025} \citep[with and without using][]{Seehafer1978} by number of volume cells.}
	\label{runtime}
\end{figure}
 
Figure \ref{runtime} shows the result of this test comparing  two variants of using \texttt{MHSXtraPy}: one for the case in which the boundary condition is not flux-balanced ("With Seehafer") and one for cases with naturally flux-balanced boundary condition ("Without Seehafer"). As the data points are approximately on a straight line in the log-log-scaled lower panel of Figure \ref{runtime} it is implied that \texttt{b3d} is approximately a polynomial time complexity algorithm in all both cases.

Due to the limitations of computational power of the laptop computer this test has been carried out on, the maximal possible discretisation $n$ was limited making this only an approximate classification of how the algorithm's runtime grows as the input size grows. While the runtime test above gives insight into the general time complexity of the algorithm and how the two variants compare to one another, its meaningfulness for the practical application to state-of-the-art line-of-sight magnetograms is limited as these typically vary significantly in size and complexity. To put this time complexity into perspective, we note that the extrapolation presented in Section \ref{sec:Example} takes on average around 12.5 seconds. 

\section{Conclusions}

\texttt{MHSXtraPy} is a newly developed code for solar magnetic field extrapolation which (i) is easy to apply due to its thorough documentation and in-code commentary, (ii) offers multiple opportunities for future development of the model and optimisation of the code, and (iii) provides an alternative to popular (NL)FF models taking into account other aspects of the physics on the Sun.

\texttt{MHSXtraPy} is a new and efficient tool for MHS extrapolation based on analytical solutions, offering a practical alternative to other methods. The code includes three analytical MHS solutions (plus options for potential and LFF extrapolation) and supports the inclusion of a non-force-free lower solar atmosphere, improving consistency with observed $\beta_p$ values. Like all analytical models, it has limitations imposed by the assumptions needed to obtain such solutions and the parameter restrictions resulting from these assumptions.

Future developments will focus on optimising the code further and exploring more models. Further validation, through comparisons with observations and results from (NL)FF or other MHS extrapolations, will be essential to assess the model's reliability and guide improvements.

\section*{Acknowledgements}

LN acknowledges financial support by the School of Mathematics and Statistics, University of St Andrews, throughout her PhD. TN acknowledges financial support by the UK's Science and Technology Facilities Council (STFC) via Consolidated Grants ST/S000402/1 and ST/W001195/1. LN thanks Jack Smith for helpful discussions and suggestions on the technical aspect of this work. 

\textit{Software}. The code and this work made extensive use of the following Python libraries/modules: NumPy \citep{numpy2020}, SciPy \citep{scipy2001, scipy2020}, Matplotlib \citep{hunter2007}, SunPy \citep{Sunpy2020, sunpy2023}, Astropy \citep{astropy2013, astropy2018, astropy2022}, and Numba \citep{numba2015}.


\section*{Data Availability}

The \texttt{MHSXtraPy} code is publicly accessible on GitHub \url{https://github.com/LMNadol/MHSXtraPy}.

All solar data used in this work were obtained from JSOC and are publicly available. The Jupyter notebook for the example presented in this paper can be found at \url{https://github.com/LMNadol/MHSXtraPy/blob/main/notebooks/paper.ipynb}. 

Bugs and feature requests may be reported through the standard GitHub procedures, or via e-mail with the corresponding author.

\section*{Conflict of Interest}

Authors declare no conflict of interest.



\bibliographystyle{rasti}
\bibliography{sorted} 





\bsp	
\label{lastpage}
\end{document}